\documentclass[sigconf, nonacm]{acmart}

\usepackage{enumitem}
\usepackage{xspace}
\usepackage[T1]{fontenc}
\usepackage{url}
\usepackage{algorithmic}
\usepackage{graphicx}
\usepackage{textcomp}
\usepackage{xcolor}

\usepackage{hyperref}
\hypersetup{hidelinks}
\usepackage{amsthm}
\usepackage{multirow}

\theoremstyle{definition}

\usepackage[inoutnumbered,linesnumbered,ruled,lined]{algorithm2e}

\SetCommentSty{mycommfont}
\SetKwInput{KwInput}{Input}                
\SetKwInput{KwOutput}{Output}              
\SetKw{Call}{Call}
\SetKw{KwPragma}{pragma}
\SetKw{KwFor}{for}
\SetKw{KwBy}{by}
\SetKw{KwOmp}{omp}
\SetKw{KwParallel}{parallel}
\SetKw{KwSchedule}{schedule}
\SetKw{KwCritical}{critical}
\SetKw{KwAtomic}{atomic}
\SetKw{KwBarrier}{barrier}
\SetKw{KwShared}{shared}
\SetKw{KwPrivate}{private}

\usepackage{caption}
\usepackage{subcaption}
\definecolor{backcolour}{rgb}{0.95,0.95,0.92}
\definecolor{bluekeywords}{rgb}{0.13,0.13,1}
\definecolor{codegray}{gray}{0.9}
\definecolor{ForestGreen}{RGB}{0,200,0}
\definecolor{mygreen}{rgb}{0,0.6,0}
\definecolor{mygray}{rgb}{0.5,0.5,0.5}
\definecolor{mymauve}{rgb}{0.58,0,0.82}
\definecolor{myteal}{RGB}{0,115,115}
\definecolor{redstrings}{rgb}{0.9,0,0}

\usepackage{listings}
\usepackage{glossaries}

\definecolor{codegray}{gray}{0.9}

\newcommand{\code}[1]{{\texttt{#1}}}

\usepackage{cleveref}

\def\allreduce{\emph{Allreduce}\xspace}
\def\api{\gls*{API}\xspace}

\def\cpu{CPU\xspace}
\def\cpus{CPUs\xspace}
\def\cuda{\gls*{CUDA}\xspace}
\def\cudaipc{\gls*{CUDA} \gls*{IPC}\xspace}

\def\d2h{\gls*{D2H}\xspace}
\def\dl{\gls*{DL}\xspace}

\def\h2d{\gls*{H2D}\xspace}

\def\hpc{\gls*{HPC}\xspace}

\def\ipc{\gls*{IPC}\xspace}

\def\llm{\gls*{LLM}\xspace}

\def\mpi{\gls*{MPI}\xspace}
\def\mpich{MPICH\xspace}

\def\nccl{\gls*{NCCL}\xspace}

\def\omb{\gls*{OMB}\xspace}

\def\openmpi{Open \gls*{MPI}\xspace}

\def\part{\emph{partitioned}\xspace}

\def\ptp{\gls*{P2P}\xspace}

\def\p2p{Peer-to-Peer\xspace}

\def\ucc{\gls*{UCC}\xspace}

\def\uct{\gls*{UCT}\xspace}
\def\ucx{\gls*{UCX}\xspace}

\def\cevent{\gls*{CUDA} \emph{event}\xspace}
\def\cevents{\gls*{CUDA} \emph{events}\xspace}

\def\cgraph{CUDA Graph\xspace}
\def\cgraphs{CUDA Graphs\xspace}
\def\ctx{\emph{context}\xspace}
\def\ctxs{\emph{contexts}\xspace}
\def\cstream{\gls*{CUDA} \emph{stream}\xspace}
\def\cstreams{\gls*{CUDA} \emph{streams}\xspace}

\def\events{\emph{events}\xspace}

\def\gpu{\gls*{GPU}\xspace}
\def\gpus{\gls*{GPU}s\xspace}
\def\gpudirect{GPUDirect\xspace}
\def\gpudirectptp{GPUDirect\_\gls*{P2P}\xspace}

\def\gpudirectrdma{GPUDirect\_\gls*{RDMA}\xspace}

\def\nvidia{NVIDIA\xspace}

\def\nvlink{NVLink\xspace}

\def\nvlinks{NVLinks\xspace}

\def\nvswitch{NVSwitch\xspace}
\def\pcie{\gls*{PCIe}\xspace}

\newacronym{AC}{AC}{Asynchronous Copy}
\newacronym{ANL}{ANL}{Argonne National Laboratory}
\newacronym{ARMCI}{ARMCI}{Aggregate Remote Memory Copy Interface}
\newacronym{API}{API}{Application Programming Interface}
\newacronym{ATS}{ATS}{Address Translation Services}
\newacronym{BLAS}{BLAS}{Basic Linear Algebra Subprograms}
\newacronym{BTB}{BTB}{Binomial Tree Based}
\newacronym{BTL}{BTL}{Byte Transfer Layer}
\newacronym{BPMF}{BPMF}{Bayesian Probabilistic Matrix Factorize}
\newacronym{CHAMPION}{CHAMPION}{Communication-aware Hardware-Assisted MPI Overlap eNgine}
\newacronym{CISC}{CISC}{Complex Instruction Set Computing}
\newacronym{CNN}{CNN}{Convolutional Neural Networks}
\newacronym{CQ}{CQ}{Completion Queue}
\newacronym{CRI}{CRI}{Communication Resources Instance}
\newacronym{CTS}{CTS}{Clear To Send}
\newacronym{cuBLAS}{cuBLAS}{NVIDIA CUDA Blas Library}
\newacronym{CUDA}{CUDA}{Compute Unified Device Architecture}
\newacronym{CUDA-Capable}{CUDA-Capable}{Capable for programming in CUDA Platform}
\newacronym{cuFFT}{cuFFT}{NVIDIA CUDA Fast Fourier Transform Library}
\newacronym{cuDNN}{cuDNN}{NVIDIA CUDA Deep Neural Networks Library}
\newacronym{cuRAND}{cuRAND}{NVIDIA CUDA Random Number Generator Library}
\newacronym{cuSPARSE}{cuSPARSE}{NVIDIA CUDA Sparse Matrix Library}
\newacronym{D2D}{D2D}{Device-to-Device}
\newacronym{DAG}{DAG}{Directed Acyclic Graph}
\newacronym{D2H}{D2H}{Device-to-Host}
\newacronym{DL}{DL}{Deep Learning}
\newacronym{DMA}{DMA}{Direct Memory Access}
\newacronym{DNN}{DNN}{Deep Neural Networks}
\newacronym{DriverAPI}{Driver API}{Driver Application Programming Interface}
\newacronym{DSL}{DSL}{Domain-Specific Language}
\newacronym{EP}{EP}{Endpoints}
\newacronym{ETL}{ETL}{Extract, Transform and Load}
\newacronym{FDS}{FDS}{Fire Dynamics Simulator}
\newacronym{FLOPS}{FLOPS}{Floating-point Operations Per Second}
\newacronym{FFT}{FFT}{Fast Fourier Transform}
\newacronym{GCC}{GCC}{GNU Compiler Collection}
\newacronym{GDR}{GDR}{GPUDirect RDMA}
\newacronym{GPU}{GPU}{Graphics Processing Unit}
\newacronym{GPGPU}{GPGPU}{General Purpose Computing on Graphics Processing Unit}
\newacronym{GSB}{GSB}{GPU Shared Buffer}
\newacronym{H2D}{H2D}{Host-to-Device}
\newacronym{H2H}{H2H}{Host-to-Host}
\newacronym{HCA}{HCA}{Host Channel Adapter}
\newacronym{HDFS}{HDFS}{Hadoop File System}
\newacronym{HMCS}{HMCS}{Hierarchical MCS}
\newacronym{HMPI}{HMPI}{Hybrid MPI}
\newacronym{HOL}{HOL}{Head-Of-Line}
\newacronym{HOOMD-blue}{HOOMD-blue}{Highly Optimized Object-oriented Many-particle
  Dynamics - Blue Edition}
\newacronym{HPC}{HPC}{High-Performance Computing}
\newacronym{HPL}{HPL}{High-Performance Linpack}
\newacronym{IB}{IB}{InfiniBand}
\newacronym{IPC}{IPC}{Inter-Process Communication}
\newacronym{ILP}{ILP}{Instruction Level Parallelism}
\newacronym{ISA}{ISA}{Instruction Set Architecture}
\newacronym{IPO}{IPO}{Input Processor Output}
\newacronym{JCUDA}{JCUDA}{Java Bindings for CUDA}
\newacronym{JNI}{JNI}{Java Native Interface}
\newacronym{JVM}{JVM}{Java Virtual Machine}
\newacronym{LAMMPS}{LAMMPS}{Large-scale Atomic/Molecular Massively Parallel Simulator}
\newacronym{LBM}{LBM}{Lattice Boltzmann Method}
\newacronym{LLM}{LLM}{Large Language Model}
\newacronym{LLN}{LLN}{Low-Level Network}
\newacronym{LMT}{LMT}{Large Message Transfer}
\newacronym{MAGC}{MAGC}{Mapping Approach for GPU Clusters}
\newacronym{MCS}{MCS}{Mellor-Crummey and Scott}
\newacronym{MIG}{MIG}{Multi-Instance GPU}
\newacronym{MIPS}{MIPS}{Million Instructions Per Second}
\newacronym{MPI}{MPI}{Message Passing Interface}
\newacronym{MPP}{MPP}{Massively Parallel Processors}
\newacronym{MPS}{MPS}{Multi-Process Service}
\newacronym{MT-MPI}{MT-MPI}{Multi-Threaded MPI}
\newacronym{Mutex}{Mutex}{Mutual Exclusion}
\newacronym{NBC}{NBC}{Non-Blocking Collective}
\newacronym{NCCL}{NCCL}{NVIDIA Collective Communications Library}
\newacronym{NIC}{NIC}{Network Interface Card}
\newacronym{NPB}{NPB}{NAS Parallel Benchmark suit}
\newacronym{NUMA}{NUMA}{Non-Uniform Memory Access}
\newacronym{NVLink-SLI}{NVLink-SLI}{NVIDIA Scalable Link Interface}
\newacronym{OMB}{OMB}{OSU Micro-Benchmarks}
\newacronym{OMPI}{OMPI}{Open MPI}
\newacronym{OPAL}{OPAL}{Open Portable Access Layer}
\newacronym{OpenACC}{OpenACC}{Open Accelerators}
\newacronym{OpenCL}{OpenCL}{Open Computing Language}
\newacronym{OpenCV}{OpenCV}{Open Computer Vision}
\newacronym{OpenGL}{OpenGL}{Open Graphic Library}
\newacronym{OpenMP}{OpenMP}{Open Multi-Processing}
\newacronym{OpenSHMEM}{OpenSHMEM}{Open Symmetric Hierarchical MEMory}
\newacronym{OS}{OS}{Operating System}
\newacronym{OSC}{OSC}{One-Sided Communication}
\newacronym{P2P}{P2P}{Point-to-Point}
\newacronym{PAP}{PAP}{Process Arrival Pattern}
\newacronym{PAT}{PAT}{Process Arrival Time}
\newacronym{PCIe}{PCIe}{Peripheral Component Interconnect Express}
\newacronym{PGAS}{PGAS}{Partitioned Global Address Space}
\newacronym{PIE}{PIE}{Position-Independent Executables}
\newacronym{POSIX}{POSIX}{Portable Operating System Interface}
\newacronym{PRQ}{PRQ}{Posted Receive Queue}
\newacronym{PRRTE}{PRRTE}{Process Management Interface (PMIx) Reference RunTime Environment}
\newacronym{Pthreads}{Pthreads}{POSIX Threads}
\newacronym{PTX}{PTX}{Parallel Thread Execution}
\newacronym{QE}{QE}{Queue Element}
\newacronym{RAW}{RAW}{Read After Write}
\newacronym{RDMA}{RDMA}{Remote Direct Memory Access}
\newacronym{RMA}{RMA}{Remote Memory Access}
\newacronym{RISC}{RISC}{Reduced Instruction Set Computing}
\newacronym{RSA}{RSA}{Reduce Scatter Allgather}
\newacronym{RTS}{RTS}{Request To Send}
\newacronym{RuntimeAPI}{Runtime API}{Runtime Application Programming Interface}
\newacronym{SHOC}{SHOC}{Scalable Heterogeneous Computing benchmark suite}
\newacronym{SHM}{SHM}{Shared Memory}
\newacronym{SIMD}{SIMD}{Single Instruction Multiple Data}
\newacronym{SIMT}{SIMT}{Single Instruction Multiple Thread}
\newacronym{SM}{SM}{Streaming Multiprocessor}
\newacronym{SMP}{SMP}{Symmetric Multi-Processor}
\newacronym{SQL}{SQL}{Structured Query Language}
\newacronym{TGCS}{TGCS}{Task Graph Computing System}
\newacronym{UCC}{UCC}{Unified Communication Collectives}
\newacronym{UCF}{UCF}{Unified Communication Framework}
\newacronym{UCM}{UCM}{Unified Communication Memory}
\newacronym{UCP}{UCP}{Unified Communication Protocols}
\newacronym{UCS}{UCS}{Unified Communication Services}
\newacronym{UCT}{UCT}{Unified Communication Transports}
\newacronym{UCX}{UCX}{Unified Communication X}
\newacronym{ULT}{ULT}{User-Level Threads}
\newacronym{UM}{UM}{Unified Memory}
\newacronym{UMQ}{UMQ}{Unexpected Message Queue}
\newacronym{UPC}{UPC}{Unified Parallel C}
\newacronym{UPI}{UPI}{Ultra Path Interconnect}
\newacronym{UVA}{UVA}{Unified Virtual Addressing}
\newacronym{VAS}{VAS}{Virtual Address Space}
\newacronym{VLIW}{VLIW}{Very Long Instruction Word}
\newacronym{VNI}{VNI}{Virtual Network Interface}
\newacronym{WAR}{WAR}{Write After Read}
\newacronym{WAW}{WAW}{Write After Write}
\newacronym{YARN}{Apache YARN}{Yet Another Resource Negotiator}

\graphicspath{{Figures/}}

\copyrightyear{2026}
\setcopyright{cc}

\begin{document}

%
\title{Accelerating Intra-Node GPU-to-GPU Communication Through Multi-Path Transfers with CUDA Graphs}

\author{Amirhossein Sojoodi}
\orcid{0000-0001-9877-3201}
\affiliation{%
  \institution{Queen's University}
  \city{Kingston}
  \state{Ontario}
  \country{Canada}
}
\email{amir.sojoodi@queensu.ca}

\author{Yıltan Hassan Temuçin}
\affiliation{%
  \institution{Queen's University}
  \city{Kingston}
  \state{Ontario}
  \country{Canada}
}
\email{yiltan.temucin@queensu.ca}

\author{Amirreza Barati Sedeh}
\affiliation{%
  \institution{Queen's University}
  \city{Kingston}
  \state{Ontario}
  \country{Canada}
}
\email{amirreza.baratisedeh@queensu.ca}

\author{Hamed Sharifian}
\affiliation{%
  \institution{Queen's University}
  \city{Kingston}
  \state{Ontario}
  \country{Canada}
}
\email{hamed.sharifian@queensu.ca}

\author{Ahmad Afsahi}
\affiliation{%
  \institution{Queen's University}
  \city{Kingston}
  \state{Ontario}
  \country{Canada}
}
\email{ahmad.afsahi@queensu.ca}

\renewcommand{\shortauthors}{Sojoodi et al.}

\begin{abstract}
Effective intra-node GPU communication is essential for optimizing performance in MPI-based HPC applications, especially when leveraging multiple communication paths. In this study, we propose a novel approach that integrates CUDA Graphs into the UCX framework to enhance intra-node multi-path point-to-point GPU communication. By concurrently leveraging multiple paths, including NVLink and PCIe through the host, and optimizing communication workflows using CUDA Graph, we achieve significant reductions in communication overhead and improve execution efficiency. To the best of our knowledge, our proposed approach is the first to seamlessly integrate CUDA Graphs into UCX. Through extensive experiments on a four-GPU node, our proposed CUDA Graph-based multi-path communication approach achieves up to a 2.95$\times$ bandwidth improvement, compared to the single-path UCX (UCT::CUDA-IPC), in GPU-to-GPU OMB bandwidth test when utilizing the host path and two other GPU paths, at message sizes up to 512MB.
\end{abstract}

%
%



\maketitle

\section{Introduction}
\label{multipath:intro}

The integration of \gpu into \hpc infrastructures has a\-t\-t\-r\-acted much attention over the past decade, coupled with the advent of many applications across various domains \cite{Bernholdt2020, top500web}. In order to efficiently utilize the computational power of \gpus, distributed applications need to allow for efficient \gpu communication. In particular, data exchange between \gpus on the same node with the help of \mpi communication protocols, the de facto standard for distributed systems \cite{mpiforumweb}, plays a critical role in reaching optimal performance.

Traditionally, \ptp communication in \gpu-a\-c\-c\-e\-l\-e\-r\-a\-t\-ed systems relies on direct data transfers through a single \nvlink or \pcie communication path, which often becomes a bottleneck and limits overall performance \cite{Temucin2021}. While several studies have attempted to improve \ptp communication by message striping across multiple paths, such as in the \ucx library, these approaches still lack the ability to provide concurrent data staging through both host and device \cite{Nukada2022,Temucin2021}. Our earlier work demonstrated that leveraging host-staged transfer alongside other \gpu-staged transfers could significantly enhance \ptp performance \cite{Temucin2021,Temuc2021b,Sojoodi2024}. Building upon this foundation, we propose a unified multi-path communication framework with two sub-designs: a static runtime-based pipelining engine and a dynamic \cgraph-based communication execution.

\cgraph enables capturing and reusing the workflow of execution that can be represented as a directed graph. Utilizing \cgraph leads to reducing the launch overhead, and has gained popularity in recent studies \cite{Choi2022a, Zheng2023}. In our proposed approach, we dynamically construct \cgraphs to encapsulate \ptp communication workflows, allowing efficient execution of data transfers with minimal launch overhead. Additionally, by caching \cgraphs in the \ucx library and reusing them for repetitive communication patterns, we improve efficiency and reduce unnecessary synchronization costs. With this approach, the \cpu is relieved from managing and orchestrating the flow, enabling more efficient data transfers between \gpus.

While our earlier work demonstrated that exploiting multiple intra-node communication paths can significantly increase \gpu-to-\gpu bandwidth \cite{Sojoodi2024}, the focus of this paper is on integrating \cgraphs directly into the \ucx transport to reduce launch and synchronization overheads. In particular, this work studies when \cgraphs provide net benefit within a multi-path communication framework and quantifies their overheads in transport-level executions. In summary, the main contributions of this paper are as follows:

\begin{enumerate}[leftmargin=0.35cm]
  \item We extend our multi-path communication framework with a \cgraph-based design that dynamically constructs and reuses communication workflows to minimize launch and intra-node communication overhead. This integration is, to the best of our knowledge, the first of its kind within the \ucx framework.
  \item We show the benefits of our proposed framework in both micro-benchmarks and a real-world application. Our experimental evaluation on a four-GPU node demonstrates that this approach achieves up to a 2.95$\times$ increase in \gpu-to-\gpu bandwidth test from OSU Micro-Benchmarks (OMB) compared to the single-path \ucx transport (UCT::CUDA-IPC). We also evaluate our framework using a distributed Jacobi application implemented with \mpi, where we achieve up to 1.28$\times$ improvement in overall execution time compared to the traditional single-path communication approach.
  \item Furthermore, we provide a comprehensive analysis of the performance trade-offs and overheads associated with \cgraphs in the context of multi-path communication, offering insights and guidelines into their practical implications for \hpc applications.
\end{enumerate}

\noindent
The remainder of this paper is organized as follows. Section \ref{multipath:bg} provides the necessary background for this work, and Section \ref{multipath:related} provides a review of the related work. Section \ref{multipath:design} describes the design and implementation of our multi-path communication framework, including both static and dynamic models. Section \ref{multipath:evaluation} presents the experimental setup, performance evaluation, and analysis of the results. Finally, Section \ref{multipath:conclusion} concludes the paper, summarizing the contributions and discussing future directions.

\section{Background and Motivation}
\label{multipath:bg}

\subsection{GPU-enabled Systems, MPI, and UCX}
\label{multipath:bg.mpiucx}

Efficient utilization of modern \gpu-enabled systems is a common challenge in \hpc frameworks and applications. These systems typically feature multiple accelerators per node, connected via high-bandwidth interconnects such as \nvlink, and to the host \cpus through \pcie. 


One of the most widely adopted communication standards in \hpc is \mpi, which provides an extensive suite of communication primitives suitable for parallel and distributed applications. Among the various implementations, \openmpi and \mpich are notable for their flexibility and widespread use in both traditional \hpc and emerging AI workloads \cite{openmpiweb,mpichweb}.

To address the evolving communication needs in applications that leverage \gpu acceleration, \mpi implementations are often reliant on high-performance communication libraries, for example, the \ucx library \cite{Shamis2015ucx}. The \ucx library is an open-source, modular communication framework intended to enable low-latency, high-bandwidth, and efficient communication through multiple hardware interconnects, including shared memory, InfiniBand, RoCE, PCIe, NVLink, etc. \cite{infinibandweb,nvidiaweb,ucxweb}.

\begin{figure}[tp]
  \centering
  \includegraphics[width=0.95\columnwidth]{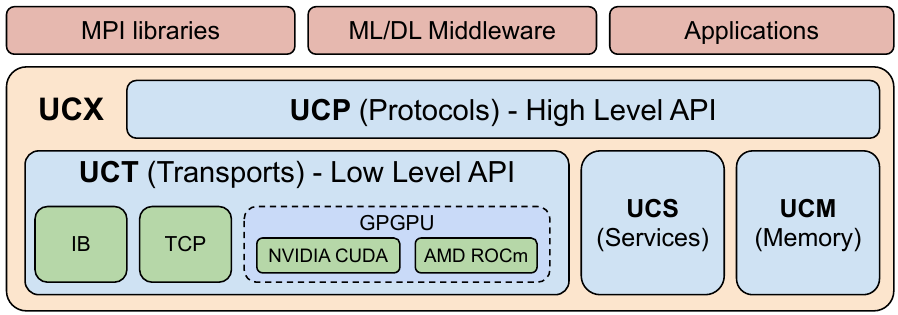}
  \caption{UCX architecture and some of its components}
  \label{multipath:figs:ucx.architecture}
\end{figure}

\Cref{multipath:figs:ucx.architecture} displays the \ucx hierarchy with a focus on important pieces, including the \uct layer. The \uct layer enables a uniform interface for communication primitives while at the same time providing hardware-specific implementations through the separate transport modules.

The \uct layer comprises several core abstractions \cite{ucxweb}: (1) \textbf{\emph{Memory Domain}} handles memory registration and memory access management specific to each hardware transport; (2) \textbf{\emph{Interface}} represents a communication resource tied to a specific transport on a given device, encapsulating the capabilities and constraints of the transport; and (3) \textbf{\emph{Endpoint}} represents a logical communication link to a remote peer, managing connection state and message flow.

\subsection{CUDA, UCX-CUDA Modules, and Direct Communication}
\label{multipath:bg.cuda}

\cuda is a widely used platform and programming model developed by \nvidia to harness the massive parallelism offered by \gpus \cite{cudaweb}. Over the past decade, several middleware solutions have emerged to integrate \gpus into distributed \hpc applications \cite{Venkata2024,Chen2023a,Sojoodi2020,Chen2022b}. While this study primarily targets \nvidia \gpus and \cuda, many of the underlying concepts are extensible to other accelerator platforms such as those from AMD and Intel.

The \ucx library offers built-in support for \cuda through its \code{uct\_cuda} transport module \cite{ucxweb}. This module consists of a shared \code{base} and three distinct submodules:
\begin{itemize} [leftmargin=0.35cm]
  \item \code{cuda\_copy}: Optimized for intra-process communication, supporting device-to-device memory transfers within the same process.
  \item \code{gdr\_copy}: Utilizes \gpudirectrdma to allow direct memory access for small messages.
  \item \code{cuda\_ipc}: Enables zero-copy inter-process communication via \cudaipc handles, making it well-suited for large messages and intra-node \mpi workloads. In our framework, we primarily target the \code{cuda\_ipc} submodule, which is critical for enabling efficient large message exchanges across \gpus residing on the same node.
\end{itemize}

Typically, when message sizes exceed 64KB, communication in \ucx defaults to a \emph{rendezvous} protocol. Depending on which process, sender or receiver, initiates the transfer and on internal configuration options, \ucx chooses between a \emph{put} or a \emph{get} operation. In both scenarios, successful communication requires that the source and destination memory buffers reside within a shared \cuda \ctx, accomplished through the exchange of \cuda \ipc memory handles.

A \cuda \ctx encapsulates a process's view of the \gpu, including memory allocations, execution resources, and synchronization primitives such as \cstreams and \cuda \events. Each \cuda \ctx is associated with a single device but can be shared across multiple threads or processes. The maintenance of a \ctx is essential to coordinate memory accesses and manage ownership.

Historically, coordinating computation and communication between \gpus required explicit staging through host memory, increasing latency and memory overhead. With the advent of \gpudirect technologies, direct communication became feasible, including \gpudirectptp, which enables direct peer-to-peer data movement between intra-node \gpus, bypassing the host.

\begin{figure}[t!]
  \centering
  \begin{subfigure}[b]{0.43\columnwidth}
    \includegraphics[width=\columnwidth]{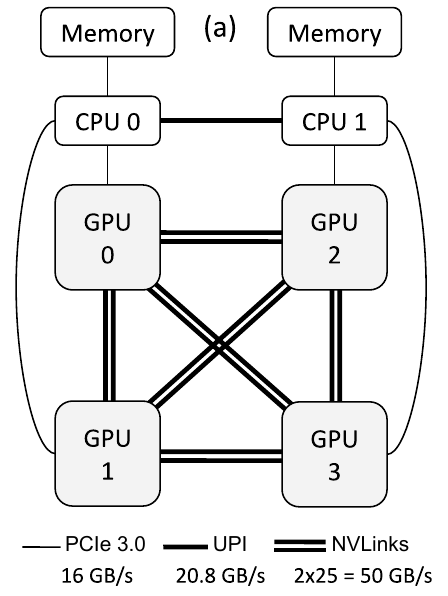}
  \end{subfigure}
  \begin{subfigure}[b]{0.438\columnwidth}
    \includegraphics[width=\columnwidth]{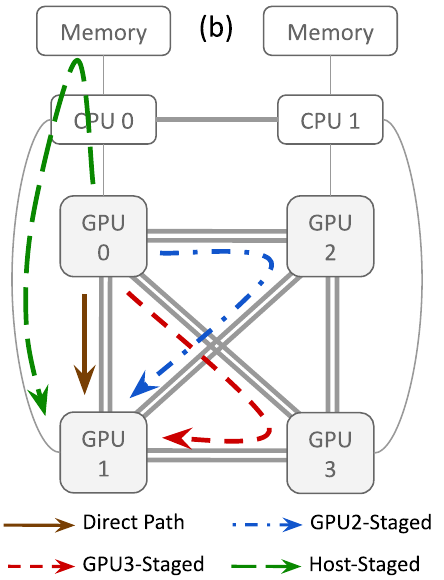}
  \end{subfigure}
  \caption{(a) A typical four-\gpu node with \nvlink (two sub-links) per \gpu pair, (b) A
    communication between \gpu-0 and \gpu-1 is split and transferred through multiple paths}
  \label{multipath:figs:node}
\end{figure}

\Cref{multipath:figs:node} illustrates both the hardware topology and a common communication scenario. In \Cref{multipath:figs:node}(a), we show a standard four-\gpu node with bidirectional \nvlink interconnects, and in \Cref{multipath:figs:node}(b), we demonstrate how a data transfer between \gpu-0 and \gpu-1 can be split across multiple paths, utilizing idle bandwidth on alternative links.

\subsection{CUDA Graphs and Advanced Runtime Integration}
\label{multipath:bg.cgraph}

Despite the potential for high bandwidth, managing concurrent communication paths introduces overhead in terms of coordination, memory management, and synchronization. As message sizes grow and communication becomes more complex, these overheads can limit scalability and performance.

To address this, we integrate \cgraph into our multi-path communication framework. \cgraph is a \cuda feature that allows users to capture a sequence of \gpu operations and replay them as a single executable graph \cite{cudagraphsweb, Choi2022a, Zheng2023}. This reduces launch overhead and enables better scheduling and optimization by the \cuda runtime. A \cgraph consists of two main components:
\begin{itemize}[leftmargin=0.35cm]
  \item \textbf{Nodes}: Represent individual \gpu operations such as memory copies or kernel launches.
  \item \textbf{Dependencies}: Specify ordering constraints between nodes, forming a directed graph.
\end{itemize}

\cgraphs can be created either via explicit \api calls or by capturing a stream of \gpu operations. By encapsulating multi-path communication as a \cgraph, we enable efficient execution of complex workflows that involve multiple memory transfers, staging buffers, and synchronization points.

Furthermore, \cgraphs can be cached and reused for repetitive patterns, reducing overhead for recurring communication tasks. This is particularly advantageous in \hpc applications with regular communication patterns, such as stencil computations or iterative solvers.

\subsection{Motivation}
\label{multipath:bg.motivation}

As illustrated in Figure \ref{multipath:figs:node}(b), traditional communication via a single path (e.g., between \gpu-0 and \gpu-1) can saturate available bandwidth, leading to performance bottlenecks. Splitting large data transfers into smaller chunks and routing them across multiple available paths can help mitigate this issue and fully exploit the system's communication capabilities.

While earlier approaches such as host-staging or manual pipelining can improve throughput, they often require careful tuning and can introduce additional \cpu overhead. Additionally, ensuring data integrity, minimizing synchronization costs, and maximizing overlap between transfers are non-trivial challenges.

Incorporating \cgraph into our design enables an automated, low-overhead solution to manage and schedule these complex workflows efficiently. It provides an elegant mechanism to orchestrate fine-grained transfers across multiple paths, thereby improving scalability and reducing \cpu involvement.

Together, these technologies form the foundation of our proposed framework: a multi-path communication model that supports both static runtime scheduling and dynamic, \cgraph-enabled execution for optimized intra-node \gpu communication in \hpc environments.

\section{Related Work}
\label{multipath:related}

\subsection{Multi-Path Communication}
\label{multipath:related:multipath}

Tatsugi and Nukada \cite{Tatsugi2022} propose a method to enhance the performance of a data transfer from a \gpu to host, by utilizing the idle \gpus. Their framework targets single-\gpu applications running on a multi-\gpu node, while our approach is designed for multi-\gpu applications.

As part of our prior studies \cite{Temucin2021,Temuc2021b}, we enhance \ptp communication within the \ucx library by utilizing host-staging multi-path communication to devise more performant collective communication for Deep Learning applications. In \cite{Sojoodi2025b}, we extend this work with a performance model to predict the optimal communication pattern for a given message size and hardware configuration, and in \cite{Sojoodi2025}, we present a heterogeneous multi-path communication framework to optimize \allreduce operations.

In \cite{Nukada2022}, Nukada proposes a method that utilizes the \pcie path to accelerate \allreduce on a multi-\gpu system. Although their method involves collectives, they follow a similar approach to our previous work to enhance each \ptp communication. Again, our approach utilizes both \nvlinks and \pcie paths, while their method is limited to \pcie.

In this work, we utilize both \nvlinks and \pcie paths to further enhance the performance of \ptp communication between intra-node \gpus, and we utilize \cgraphs inside the \ucx library to optimize the communication workflow.

\subsection{CUDA Graphs}
\label{multipath:related:cgraph}

Numerous studies have demonstrated that \cgraphs effectively reduce \cpu overhead by consolidating multiple kernel launches into a single execution unit, thereby enhancing \gpu utilization and overall application performance \cite{Zheng2023, Zhao2023, Lin2021, Ekelund2025, Qiao2020}. For example, frameworks have reported substantial speedups, such as up to 3.47$\times$ in kernel execution \cite{Zhao2023} and over 1.4$\times$ through kernel batching in iterative \hpc applications \cite{Ekelund2025}. Beyond reducing launch overhead, \cgraphs also provide the opportunity for more efficient resource management, including notable reductions in \gpu memory consumption via metadata optimizations \cite{Zheng2023}. These improvements underscore the role of \cgraphs in achieving high performance in \hpc workloads.

To facilitate the adoption of this feature, several frameworks and programming models have emerged that simplify \cgraph development \cite{Huang2021, Lin2021, Qiao2020}. Lin et al.\cite{Lin2021} proposed a lightweight programming framework that abstracts the complexity of constructing \cgraphs for \gpu computation. Similarly, Huang et al.\cite{Huang2021} introduced a task-graph programming model designed to aid the development of \gls*{TGCS}, where function calls and their inter-dependencies are represented as task graphs; these are then submitted using \cgraphs on supported \gpus. Qiao et al. \cite{Qiao2020} leveraged \cgraphs within a source-to-source compiler for image processing \glspl*{DSL}, achieving improved workflow optimization through graph-based execution.

Libraries like \nccl and \ucc have also integrated \cgraph support to optimize collective communication operations \cite{ncclweb,uccweb}. \nccl utilizes \cgraphs to capture and replay collective operations, reducing overhead, and \ucc uses \cgraphs to group multiple data transfers between intra-node \gpus. At the time of writing this paper, \ucc does not put computations on the \cgraphs, and it only uses them for communication operations. Our work is different from both of these libraries, as we focus on optimizing \ptp communication between intra-node \gpus by leveraging multiple paths (with staging) and utilizing \cgraphs to encapsulate the entire communication workflow.

Despite these advances, challenges remain in fully exploiting \cgraphs, particularly for dynamic workloads and complex control flows \cite{dynamiccudagraphsweb}. Many implementations struggle with applications featuring varying input sizes or data-dependent execution paths, often requiring intricate compiler or runtime support and incurring higher memory overhead \cite{Zheng2023}. Additionally, optimizing dynamic and irregular communication/computation patterns using \cgraphs paradigm continues to be challenging. Addressing these limitations is essential for extending the applicability of \cgraphs to a broader spectrum of modern \hpc and scientific computing workloads.

\subsection{Comparison with our Previous Work}
\label{multipath:related:comparison}

In \cite{Sojoodi2024}, we introduced a multi-path communication framework within the \ucx library to enable multi-path \ptp communication between intra-node \gpus through both \nvlink and \pcie interconnects. In the current work, we extend this framework further by integrating \cgraphs to optimize communication workflows and reduce synchronization costs. Our approach allows for dynamic construction and caching of \cgraphs to save execution time and reduce overheads associated with launching multiple \gpu operations. We build this \cgraph engine inside the \ucx library, to provide a seamless and unified interface for users to leverage \cgraphs in their applications.  Furthermore, we provide a comprehensive performance evaluation of \cgraphs lifecycle in the context of multi-path communication, analyzing their overheads and benefits when applied to middleware libraries like \ucx.

Our work is distinct from collective communication libraries such as \nccl and \ucc. While these libraries have incorporated \cgraphs to optimize the \textbf{collective operations}, our work integrates \cgraphs directly into the \ucx transport layer to accelerate \ptp communication. This enables graph-based execution of fine-grained, staged, multi-path \ptp transfers that are transparent to upper-layer libraries and applications, including collectives built on top of \ucx.

\section{Design and Implementation}
\label{multipath:design}


To enable efficient and scalable communication between \gpus in the same node, we have designed a system that builds upon the existing functionality of \code{uct\_cuda} transport module in \ucx. This is built by adding both a multi-path communication engine, which supports the use of multiple communication paths simultaneously, and a runtime acceleration mechanism based on \cgraph, which minimizes the synchronization costs. Our overall design philosophy is to prioritize maintaining compatibility with the internal structure of \ucx, introducing innovative features on top of it, and utilizing the advantages of concurrent paths, with minimal overhead additions. In this section, we will present a detailed description of our architecture, describe the pipelining mechanism that we have used, discuss several strategies for runtime tuning, and ensure that we discuss the guarantees pertaining to data integrity in detail.

\subsection{Design Objectives}

Our design addresses the following core goals to support high-performance communication:

\begin{enumerate} [leftmargin=0.35cm]
  \item \textbf{Multi-GPU awareness:} Automatically detect and utilize all available \gpus and their interconnects for efficient communication path construction within \ucx instances.
  \item \textbf{Path selection:} Dynamically select the most suitable and available paths between source and destination \gpus, including direct and indirect routes via staging \gpus or host.
  \item \textbf{Communication scheduling:} Efficiently distribute data along the selected paths using fine-grained pipelining.
  \item \textbf{Path optimization:} Maximize throughput and minimize idle time by overlapping transfers across paths.
  \item \textbf{Data integrity:} Maintain correctness by ensuring ordered, synchronized, and contention-free transfers.
  \item \textbf{Low overhead:} Keep the framework's runtime and memory overhead negligible compared to the overall communication cost.
  \item \textbf{Seamless integration:} Ensure compatibility with existing \ucx applications, requiring minimal changes to leverage the new capabilities.
  \item \textbf{Dynamic adaptability:} Support dynamic graph construction and caching to optimize repetitive communication patterns.
\end{enumerate}

\subsection{Framework Architecture}
\label{multipath:design.framework}

Figure \ref{cgraph:figs:design} presents a simplified overview of our enhanced framework. The major components are described below:

\begin{figure}[tp]
  \centering
  \includegraphics[width=0.95\columnwidth]{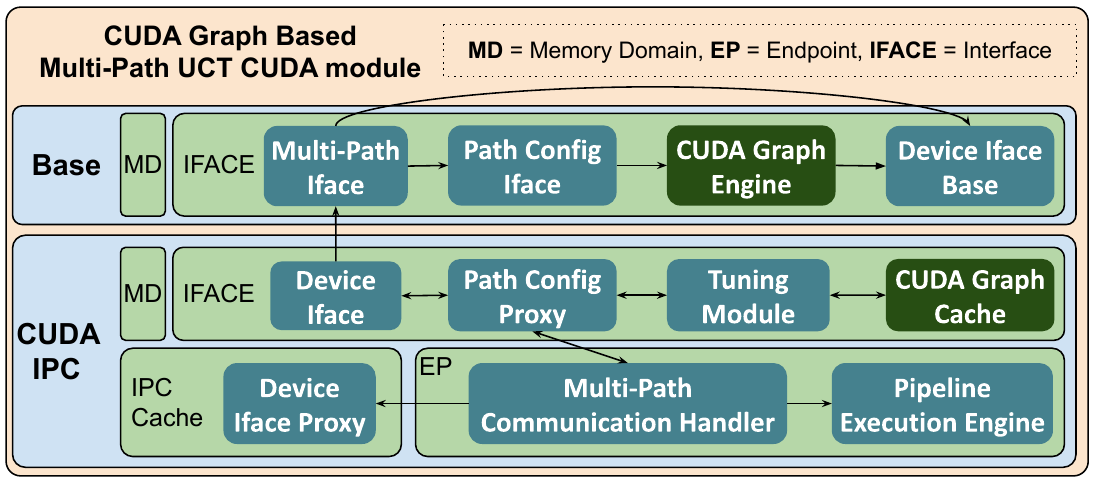}
  \caption{Interaction between the Multi-Path Communication Handler, the 2-D Pipelining Engine, and the CUDA Graph Engine. The handler selects the communication paths, and the pipelining engine determines chunking and ordering. Then these configurations are passed to the CUDA Graph Engine, which constructs the graph that executes the staged multi-path workflow. The graph is cached for future reuse.}
  \label{cgraph:figs:design}
\end{figure}

\begin{itemize}[leftmargin=0.35cm]
  \item \textbf{Base Module:} This module is responsible for probing the system's hardware topology and initializing internal structures for the framework. It detects the available \gpus and their interconnects, and creates the necessary runtime resources for each \gpu, including \cstreams, \cevents, device buffers, path metadata, and the communication proxy objects that bridge \cuda \ipc and the graph engine. Some of these functionalities are inherited from the original \code{uct\_cuda} module, and we use the already available functionalities to detect the \gpus and their interconnects. However, the rest of the explained functionalities are newly implemented as part of our framework.
  \item \textbf{CUDA IPC Module:} This module handles incoming communication requests, selecting the suitable paths, handling configurations from environment variables, and scheduling the communication along the selected paths with the 2-D pipelining engine. The connection between the \code{base} and \code{cuda\_ipc} modules is established through proxy entities.
  \item \textbf{CUDA Graph Engine:} This engine constructs reusable \cgraphs in a lazy fashion using the explicit \cuda Driver \api. It abstracts the pipelined multi-path workflow into a single \cgraph, reducing launch overhead and synchronizations. Each graph represents a unique communication configuration between two \gpus, with varying numbers of paths and chunks.
  \item \textbf{CUDA Graph Cache:} The cache stores instantiated \cgraphs based on the communication configuration, including source/destination buffer addresses and message size. It uses a configurable least-recently-used (LRU) eviction policy and is implemented as a fixed-size hash table, tunable via environment variables.
\end{itemize}


\subsection{2-D Pipelining Engine and CUDA Graph Execution}
\label{multipath:design.2d}

\begin{figure}[!t]
  \centering
  \includegraphics[width=\columnwidth]{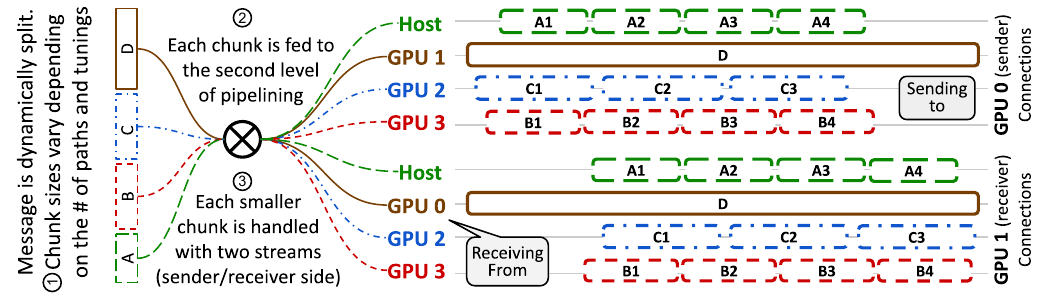}
  \caption{A simplified view of 2-D pipelined communication from \gpu-0 to \gpu-1 using the available \nvlinks and \pcie. Staging GPUs' timelines are not shown for simplicity.}
  \label{multipath:figs:transfer.pipeline}
\end{figure}

\begin{figure}[tp]
  \centering
  \includegraphics[width=0.95\columnwidth]{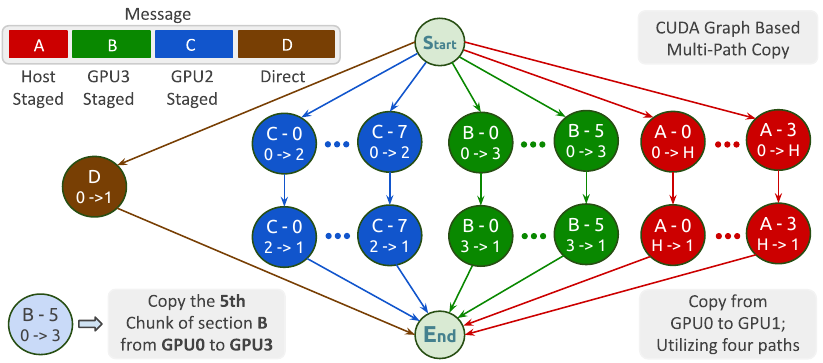}
  \caption{A \cgraph-based multi-path communication from \gpu-0 to \gpu-1 using the available \nvlinks and \pcie. The graph in this figure is generated by the CUDA Graph Engine shown in Figure \ref{cgraph:figs:design}. Each node denotes a memory-copy operation, and edges represent the control-flow dependencies ensuring staged transfers execute in the required order.}
  \label{cgraph:figs:transfer.graph}
\end{figure}

Our 2-D pipelining engine is the core of the data distribution mechanism. As shown in Figure \ref{multipath:figs:transfer.pipeline}, a large message is first partitioned according to the number of selected paths (horizontal split). Then, each path is further subdivided into smaller chunks (vertical split) that are transferred in a pipelined fashion.

The number of chunks per path is a tunable parameter depending on message size and the capabilities of each link. Our experiments indicate that a chunk size of 1MB generally provides a good balance between overhead and concurrency. However, to provide maximum overlap between paths, we use the tuned chunk count as described in \Cref{multipath:design.tuning}. 

Each data chunk is processed using two separate \cstreams: one for the source-to-staging transfer and another for the staging-to-destination transfer. This parallelization allows communication along multiple routes to proceed concurrently, while also enabling fine-grained overlap within each route.

When \cgraphs are enabled, the entire communication workflow, including chunk partitioning, staging transfers, synchronization, and final commit, is encapsulated into a Directed Acyclic Graph (DAG), as shown in Figure \ref{cgraph:figs:transfer.graph}. Each node in the DAG represents a distinct memory copy or kernel operation, and edges represent control-flow or execution dependencies. This abstraction reduces per-transfer launch overhead, improves reuse for repetitive patterns, and allows for better optimization by the \cuda runtime scheduler.

In the presence of \cgraphs, the multi-path workflow is orchestrated as follows: the Multi-Path Communication Handler selects the available paths and delegates to the 2-D Pipelining Engine, which determines chunk distribution and ordering. This engine then issues a structured workflow to the \cgraph Engine, which builds or retrieves the required \cgraph for execution. \Cref{cgraph:alg:ucx_cudagraph_transfer} and \Cref{cgraph:alg:ucx_cudagraph_functions} outline the algorithms for creating and launching these \cgraphs. As shown in \Cref{cgraph:alg:ucx_cudagraph_transfer}, the \code{CreateGraph} function is responsible for constructing a new \cgraph instance based on the communication parameters, while the \code{LaunchGraph} function retrieves and launches a \cgraph instance.

\begin{algorithm}[!t]
  \small
  \SetInd{0.2em}{0.8em}
  \DontPrintSemicolon

  \SetKwProg{Fn}{Function}{:}{end}
  \SetKwFunction{CreateGraph}{CreateGraph}
  \SetKwFunction{LaunchGraph}{LaunchGraph}
  \SetKwFunction{GetPathConfig}{GetPathConfig}
  \SetKwFunction{UpdateTempIDs}{UpdateTempIDs}
  \SetKwFunction{OffsetPtr}{OffsetPtr}
  \SetKwFunction{PeerToPeerCopy}{PeerToPeerCopy}
  \SetKwFunction{StageHostCopy}{StageHostCopy}
  \SetKwFunction{StageGPUCopy}{StageGPUCopy}

  \Fn{\CreateGraph{$src, dst, size, ctx_{src}, ctx_{dst}, stream$}}{
    $id_{src}, id_{dst} \leftarrow$ get device IDs from $ctx_{src}, ctx_{dst}$ \;
    $graph\_id \leftarrow$ get unique graph ID based on $src, dst, size$ \;
    $config \leftarrow$ \GetPathConfig{$size$} \;
    \UpdateTempIDs{$config, id_{src}, id_{dst}$} \;
    $sent\_data \leftarrow 0$ \;
    \ForEach{path $p$ in $config$}{
      $chunk\_size[p] \leftarrow \lceil (size \times share[p]) / max\_chunks \rceil$ \;
      $src_p \leftarrow$ \OffsetPtr{$src, sent\_data$} \;
      $dst_p \leftarrow$ \OffsetPtr{$dst, sent\_data$} \;
      $sent\_data \mathrel{+}= chunk\_size[p]$ \;
      \If{$config.tmp\_dev[p]$ is direct}{
        \PeerToPeerCopy{$src_p, dst_p, chunk\_size[p],$ \newline \hspace*{2em} $id_{src}, id_{dst}, streams[p], graph$} \;
      }
      \ElseIf{$config.tmp\_dev[p]$ is host}{
        \StageHostCopy{$src_p, dst_p, chunk\_size[p],$ \newline \hspace*{2em} $id_{src}, id_{dst}, streams[p], graph$} \;
      }
      \ElseIf{$config.tmp\_dev[p]$ is GPU}{
        \StageGPUCopy{$src_p, dst_p, chunk\_size[p],$ \newline \hspace*{2em} $id_{src}, id_{dst}, config.tmp\_dev[p], streams[p], graph$} \;
      }
    }
    store constructed graph instance with $graph\_id$ in cache \;
  }

  \Fn{\LaunchGraph{$ctx_{src}, ctx_{dst}, stream, graph\_id$}}{
    $instance \leftarrow$ get graph instance with $graph\_id$ from cache \;
    \If{$instance$ is NULL}{
      $instance \leftarrow$ \CreateGraph{$src, dst, size, ctx_{src}, ctx_{dst}, stream$} \;
    }
    launch $instance$ on appropriate stream \;
    record event and synchronize with input $stream$ \;
  }

  \caption{CUDA Graph Creation and Launch for Multi-Path UCX Transfers}
  \label{cgraph:alg:ucx_cudagraph_transfer}
\end{algorithm}

\begin{algorithm}[!t]
  \small
  \SetInd{0.2em}{0.8em}
  \DontPrintSemicolon

  \SetKwProg{Fn}{Function}{:}{end}
  \SetKwFunction{PeerToPeerCopy}{PeerToPeerCopy}
  \SetKwFunction{StageHostCopy}{StageHostCopy}
  \SetKwFunction{StageGPUCopy}{StageGPUCopy}

  \Fn{\PeerToPeerCopy{$src, dst, size, id_{src}, id_{dst}, streams, graph$}}{
    \For{$s \leftarrow 0$ \KwTo $streams - 1$}{
      create memcpy node: device $\rightarrow$ device \;
      config node with $size$ and device ids ($id_{src}, id_{dst}$) \;
      add memcpy node from $src[s]$ to $dst[s]$ on $graph$ \;
    }
  }

  \Fn{\StageHostCopy{$src, dst, size, id_{src}, id_{dst}, streams, graph$}}{
    \For{$s \leftarrow 0$ \KwTo $streams - 1$}{
      create memcpy node: device ($id_{src}$) $\rightarrow$ host (staging) \;
      create memcpy node: host $\rightarrow$ device ($id_{dst}$) \;
      add dependency between staging and completion nodes \;
    }
  }

  \Fn{\StageGPUCopy{$src, dst, size, id_{src}, id_{dst}, id_{stage}, streams, graph$}}{
    \For{$s \leftarrow 0$ \KwTo $streams - 1$}{
      create memcpy node: device ($id_{src}$) $\rightarrow$ staging GPU ($id_{stage}$) \;
      create memcpy node: staging GPU ($id_{stage}$) $\rightarrow$ device ($id_{dst}$) \;
      add dependency between staging and completion nodes \;
    }
  }

  \caption{CUDA Graph Nodes Creation and Configuration}
  \label{cgraph:alg:ucx_cudagraph_functions}
\end{algorithm}

In more detail, \Cref{cgraph:alg:ucx_cudagraph_transfer} starts by determining the source and destination device IDs from their respective \cuda \ctxs (Line 4). It then generates a unique graph ID based on the source/destination addresses and message size (Line 5). The path configuration is obtained using the \code{GetPathConfig} function, which considers factors such as message size and available paths (Line 6). Temporary device IDs for staging are updated according to the source and destination devices and the configuration (Line 7). The message is then partitioned into chunks for each path, and the appropriate copy functions are invoked based on whether the path is direct or staged (Lines 9-21). In our original implementation, we used a round-robin approach to distribute chunks across paths one-by-one, but we provide the simplified version here for clarity. Finally, the constructed graph instance is stored for future use (Line 22).

In the \code{LaunchGraph} function, the graph instance is retrieved using the unique graph ID (Line 28) from the cache. If the instance does not exist, it is created using the \code{CreateGraph} function (Lines 29-31). The graph is then launched on the appropriate stream (Line 32), and a \cevent is recorded on the \cstream and passed to \ucx for further processing (Line 33).

In \Cref{cgraph:alg:ucx_cudagraph_functions}, we detail the helper functions used to create specific types of memory copy nodes, including direct peer-to-peer copies and staged transfers via host or intermediate \gpus. In each function, we create the necessary \cgraph nodes and configure them with the appropriate parameters, such as source/destination addresses, sizes, and device IDs. These nodes are then added to the graph instance passed by reference. In Lines 18 and 29, we add the necessary dependencies to ensure the correct ordering of operations for staged transfers of chunks.

\subsection{Runtime Tuning and Configuration}
\label{multipath:design.tuning}

To accommodate diverse hardware topologies and workloads, our framework supports both user-defined and auto-tuned configurations.

\begin{itemize}[leftmargin=0.35cm]
  \item \textbf{Environment Configuration:} Users can configure the framework via environment variables to enable or disable specific paths (e.g., host-staging), specify the number of concurrent paths, and define the number of chunks per path. These configurations allow flexibility for both expert users and general-purpose deployment.
  \item \textbf{\cgraphs based Topology Tuning:} We extend the offline tuning mechanism to also support \cgraph-based workflows. This tuning stage exhaustively searches for optimal communication configurations based on the number of \gpus, available links with various bandwidths and latencies, and transfer sizes. The tuned parameters are used as defaults unless overridden by the user. Note that with the \cgraph integration, this tuning should be performed separately for \cgraph-enabled and non-\cgraph modes, as the optimal configurations may differ due to the reduced overheads and improved scheduling capabilities of \cgraphs. Therefore, the number of paths and the number of chunks per path are fixed values determined by the offline topology tuning mechanism. These configurations are used in all evaluation experiments to ensure optimal performance.
\end{itemize}

\subsection{Ensuring Data Integrity and Correctness}
\label{multipath:design.integrity}

To preserve correctness, our design ensures strict data consistency and avoids race conditions by leveraging \cuda's synchronization semantics. The following guarantees are enforced with both the non-\cgraph and \cgraph-based implementations:

\begin{itemize}[leftmargin=0.35cm]
  \item \textbf{Contention Avoidance:} For each directional path (e.g., \gpu-2 to \gpu-3), only one transfer is allowed at a time. This is achieved by mapping all transfers for a specific direction to the same \cstream and synchronizing using \cevents.
  \item \textbf{Dependency Handling:} For paths involving staging buffers (e.g., via host or intermediate \gpu), the second transfer (staging to destination) is issued only after the first completes.
  \item \textbf{Ordering Guarantee:} Transfers for each chunk within a message are scheduled in a well-defined order and completed before subsequent messages are handled. Reordering is not an issue in our design because each chunk writes into a pre-allocated, non-overlapping region of the destination buffer. The sender determines the destination offset for every chunk before initiating the transfer. Final completion is signaled only after all chunk-specific synchronization points indicate successful arrival.
  \item \textbf{Final Synchronization:} All \cstreams are synchronized at the end of a message to ensure that the transfer has completed and that the data in the destination buffer is consistent and ready for use. As explained earlier, this is done by recording a \cevent on each stream within each path and instructing the main \cstream to wait for all events to complete before returning control to the caller. This is done in a non-blocking manner, allowing \ucx to continue processing other requests while waiting for the communication to complete.
\end{itemize}

These mechanisms ensure that pipelined and concurrent communications do not result in corrupted or out-of-order data, making the framework suitable for latency-sensitive and bandwidth-bound applications.

\section{Evaluation}
\label{multipath:evaluation}

To assess the performance benefits and practical effectiveness of our proposed framework, we evaluated both the pipelined multi-path communication engine and the \cgraph-based runtime extensions under a diverse set of benchmarks. These include synthetic micro-benchmarks as well as an application-level use case. The micro-benchmarks consist of \ucx Put Bandwidth, \omb \mpi Bandwidth (\code{OMB\_BW}), and \omb \mpi Bidirectional Bandwidth (\code{OMB\_BIBW}) \cite{Bureddy2012}. For application-level evaluation, we used the Jacobi iterative solver \cite{jacobiweb}, a representative stencil computation pattern.

Moreover, to analyze the behavior and impact of various configurations of \cgraphs in our multi-path framework, we performed a series of experiments using \omb \mpi \code{ Latency} tests. These tests help us understand how different graph structures and sizes affect performance, particularly in terms of latency reduction and overhead minimization in various stages of a \cgraph lifecycle, including creation, construction, instantiation, and launch.

\subsection{Experimental Setup}

All experiments were conducted on two multi-\gpu compute nodes from the Digital Research Alliance of Canada clusters: Beluga and Narval. The Beluga node is equipped with four \nvidia V100 \gpus, each connected via two bidirectional \nvlink links in a full-mesh topology. The Narval node includes four \nvidia A100 \gpus with a denser interconnect, where each pair of \gpus is connected by four \nvlinks. The full topological layout of Beluga is depicted in Figure \ref{multipath:figs:node}. 

Both systems ran \ucx version 1.14.0. For \mpi support, we used \openmpi version 5.0.4 for both the non-\cgraph and \cgraph-based evaluations. In all tests, unless otherwise noted, we used pinned host memory and \cuda \ipc for all device memory allocations and transfers. We also performed these experiments 1000 times and report the average results. All performance comparisons in this section use the traditional single-path UCX (UCT::CUDA-IPC) transport as the baseline.

\subsection{UCX Put Bandwidth Evaluation}

Figure \ref{multipath:figs:figure_ucx_put_bw} shows the performance of \ucx Put operations measured with two ranks, one on each \gpu, on both Beluga (top) and Narval (bottom) clusters. We compare the baseline performance of default \ucx (\uct::CUDA-IPC) against our pipelined multi-path and \cgraph-based framework. The key observations include:

\begin{figure}[t!]
  \centering
  \includegraphics[width=0.85\columnwidth]{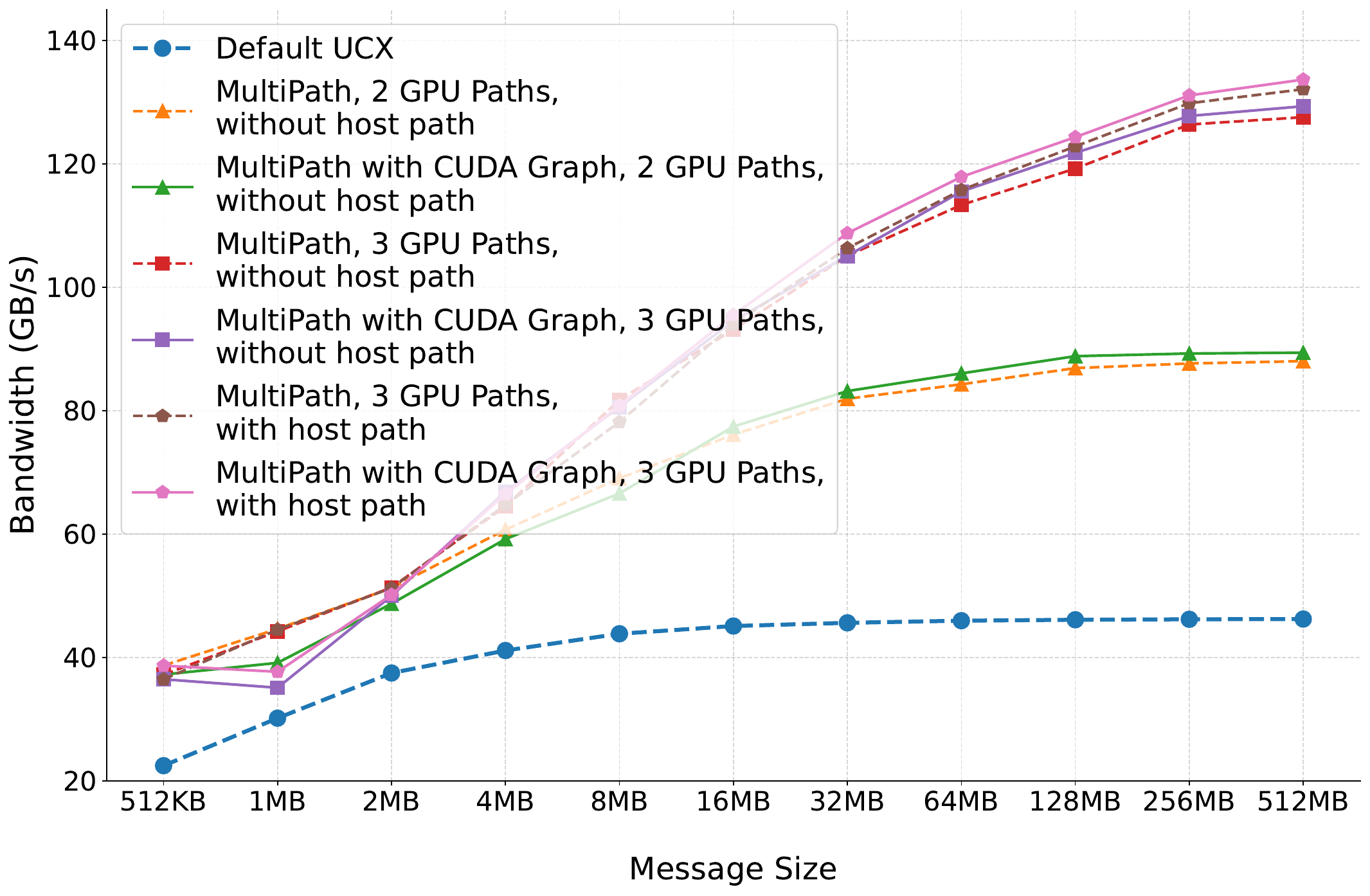}
  \includegraphics[width=0.85\columnwidth]{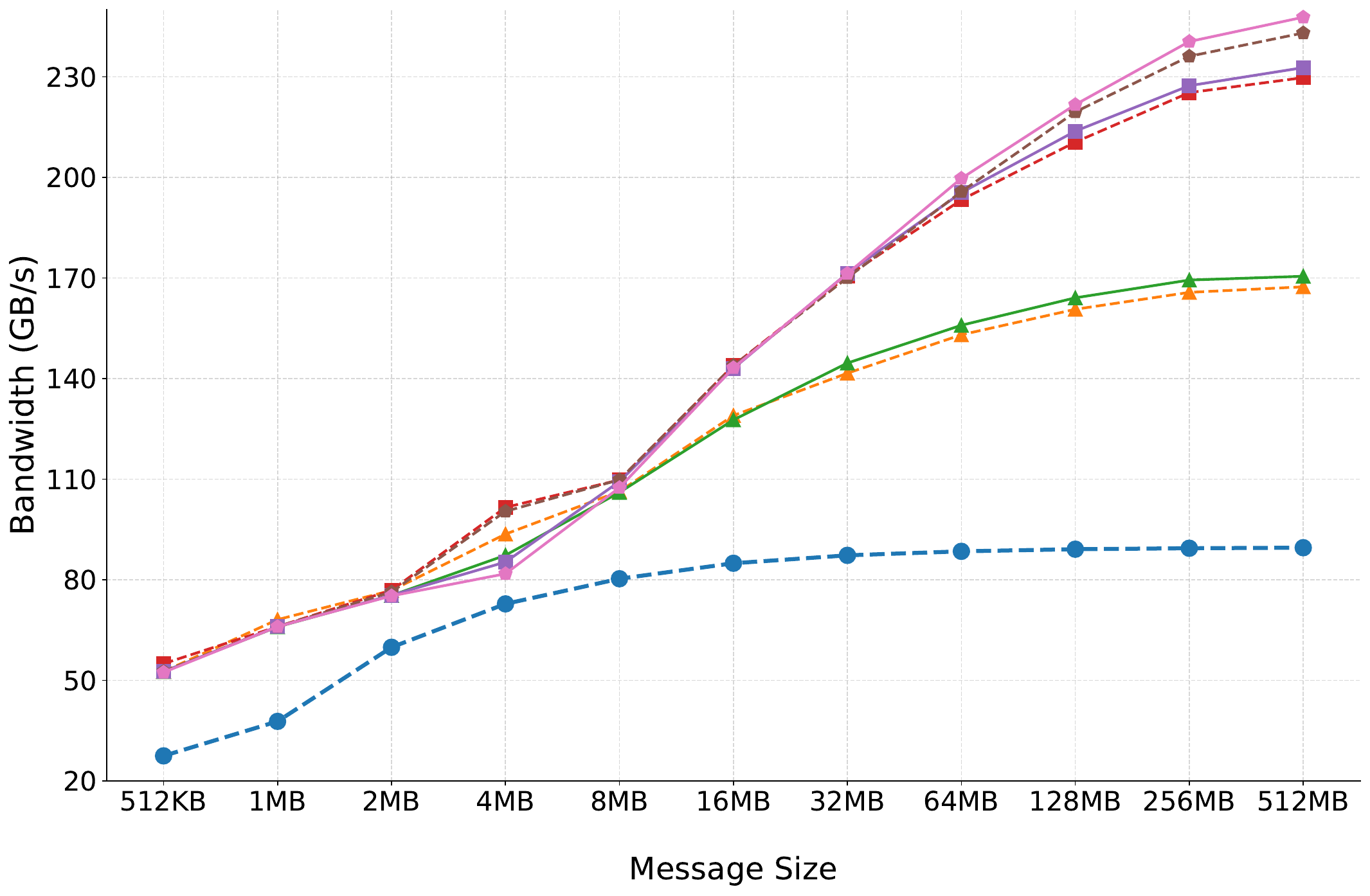}
  \caption{\ucx Put Bandwidth comparison of our multi-path framework with and without \cgraph against default \ucx (\uct::CUDA-IPC), on Beluga (top) and Narval (bottom)}
  \label{multipath:figs:figure_ucx_put_bw}
\end{figure}

\begin{enumerate} [leftmargin=0.35cm]
  \item When using three \gpu paths and optionally including a host-staging path, our approach achieves up to 2.85$\times$ and 2.75$\times$ bandwidth improvements over the baseline on Beluga and Narval, respectively, for message sizes larger than 32MB, which is important for bandwidth-bound applications such as \llm and \dl workloads.
  \item The integration of \cgraphs further enhances performance, especially for larger data sizes, due to the increased number of nodes/operations in the \cgraph comparing to the small data sizes. The results show up to 2.95$\times$ and 2.85$\times$ speedup on Beluga and Narval, respectively, for the same size range. The reason for more improvement for larger message sizes is due to the fact that the number of operations in the \cgraph is larger. For instance, for a 128MB message size, where we use three \gpu paths and four chunks per path (on average) the created \cgraph has 16 memory copy nodes, with four chunks per path on average. Packing more operations into a single \cgraph reduces the launch overhead and synchronization costs on the \cpu side, leading to better performance.
  \item Including the host path contributes marginally to the bandwidth improvement (up to 15\%), which is expected given the lower performance of \pcie relative to \nvlink. In many scenarios, this contribution is negligible compared to enabling additional direct \gpu paths.
  \item To avoid clutter in the figure, intermediate configurations (e.g., one or two \gpu paths with host path enabled) are omitted, since they show similar patterns to the configurations with three \gpu paths.
\end{enumerate}

\subsection{MPI Micro-benchmark Results}
\label{multipath:results.omb}

\begin{figure}[t!]
  \centering
  \includegraphics[width=0.99\columnwidth]{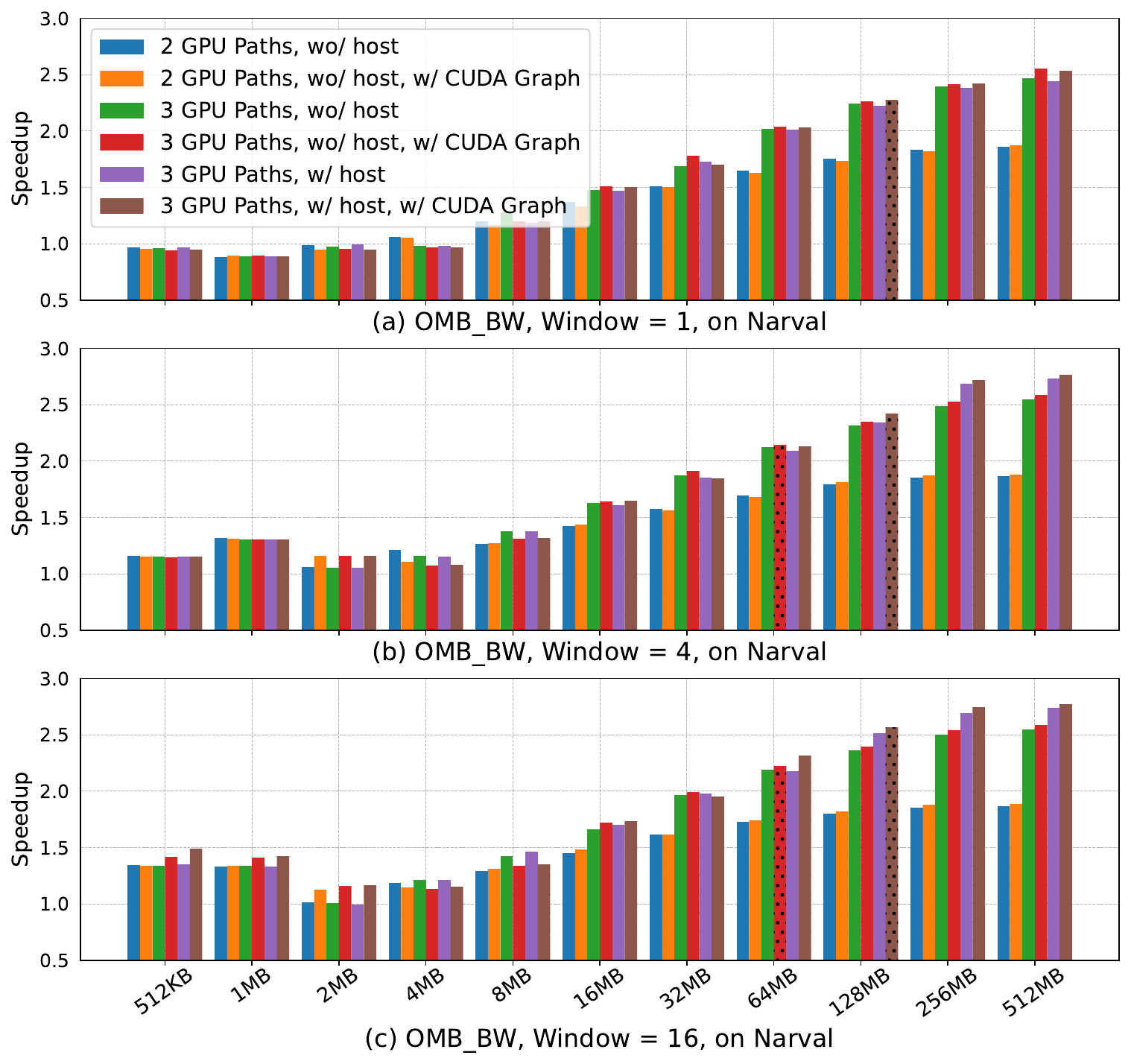}
  \caption{Multi-Path OMB Unidirectional MPI Bandwidth (BW) comparison with and without CUDA Graph, on Narval}
  \label{cgraph:figs:omb_bw_narval}
\end{figure}

\begin{figure}[t!]
  \centering
  \includegraphics[width=0.99\columnwidth]{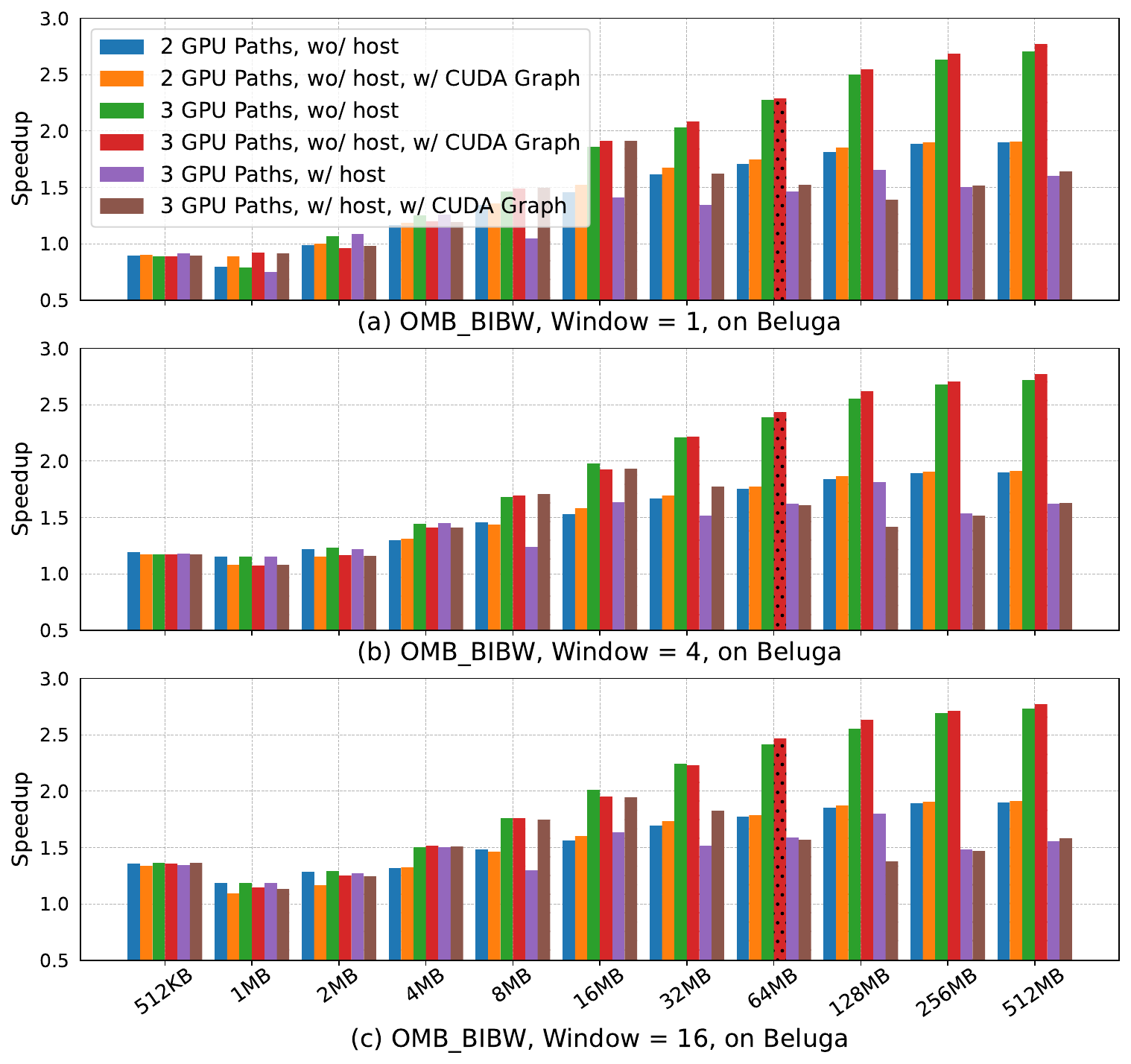}
  \caption{Multi-Path OMB Bidirectional MPI Bandwidth (BIBW) comparison with and without CUDA Graph, on Beluga}
  \label{cgraph:figs:omb_bibw_beluga}
\end{figure}

\begin{figure}[t!]
  \centering
  \includegraphics[width=0.99\columnwidth]{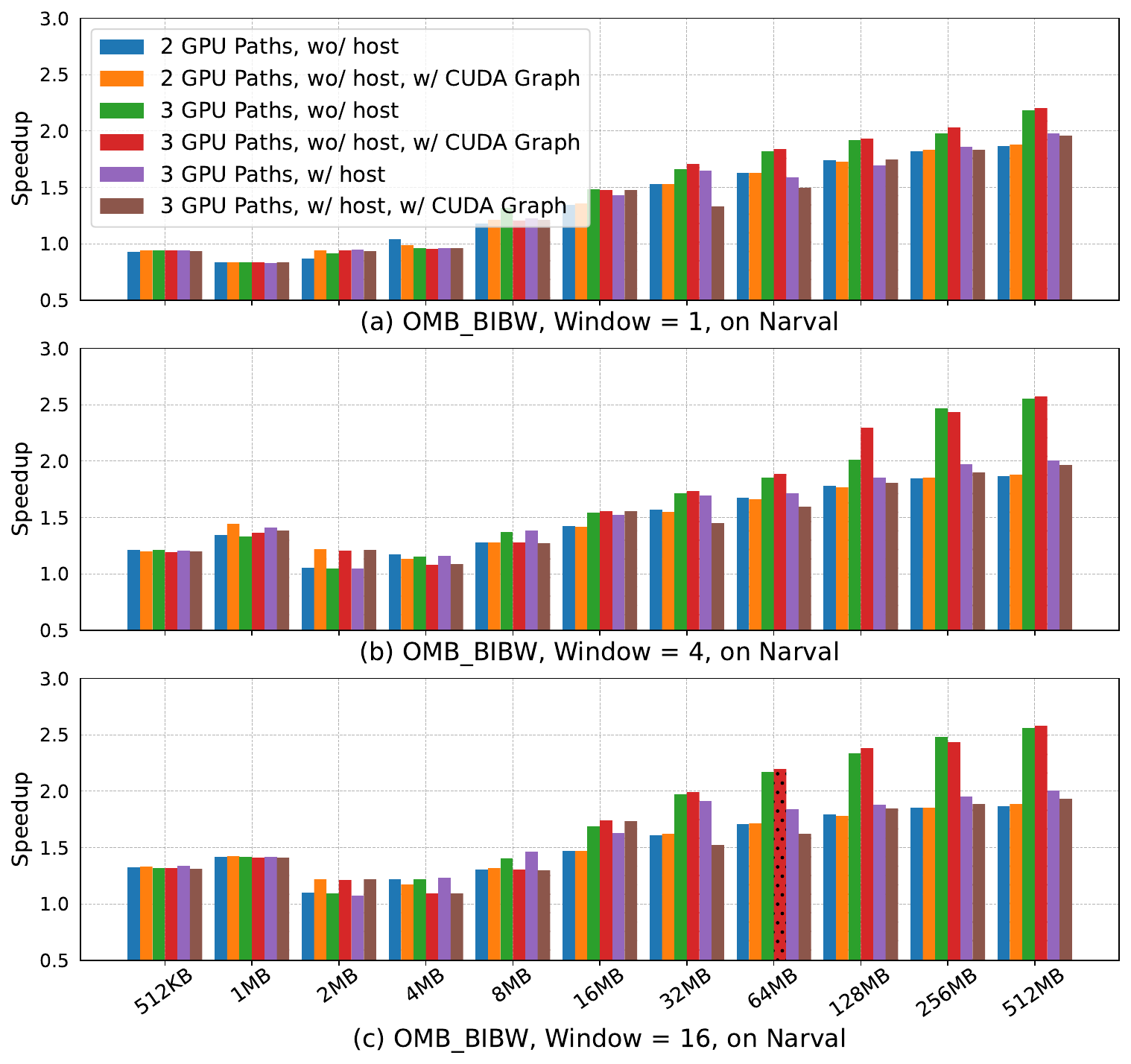}
  \caption{Multi-Path OMB Bidirectional MPI Bandwidth (BIBW) comparison with and without CUDA Graph, on Narval}
  \label{cgraph:figs:omb_bibw_narval}
\end{figure}

Figure \ref{cgraph:figs:omb_bw_beluga} and Figure \ref{cgraph:figs:omb_bw_narval} present \omb \mpi unidirectional bandwidth (BW) measurements on Beluga and Narval, respectively. Also, Figure \ref{cgraph:figs:omb_bibw_beluga} and Figure \ref{cgraph:figs:omb_bibw_narval} depict the bidirectional bandwidth (BIBW) tests for both clusters. These tests were conducted with two \mpi ranks, each assigned to one \gpu, and various message sizes ranging from 1MB to 64MB. All tests were conducted across different message sizes and \emph{window} sizes (1, 4, and 16). The \emph{window} size defines how many messages can be posted without waiting for prior ones to complete. We observe the following key results:

\begin{figure}[t!]
  \centering
  \includegraphics[width=0.99\columnwidth]{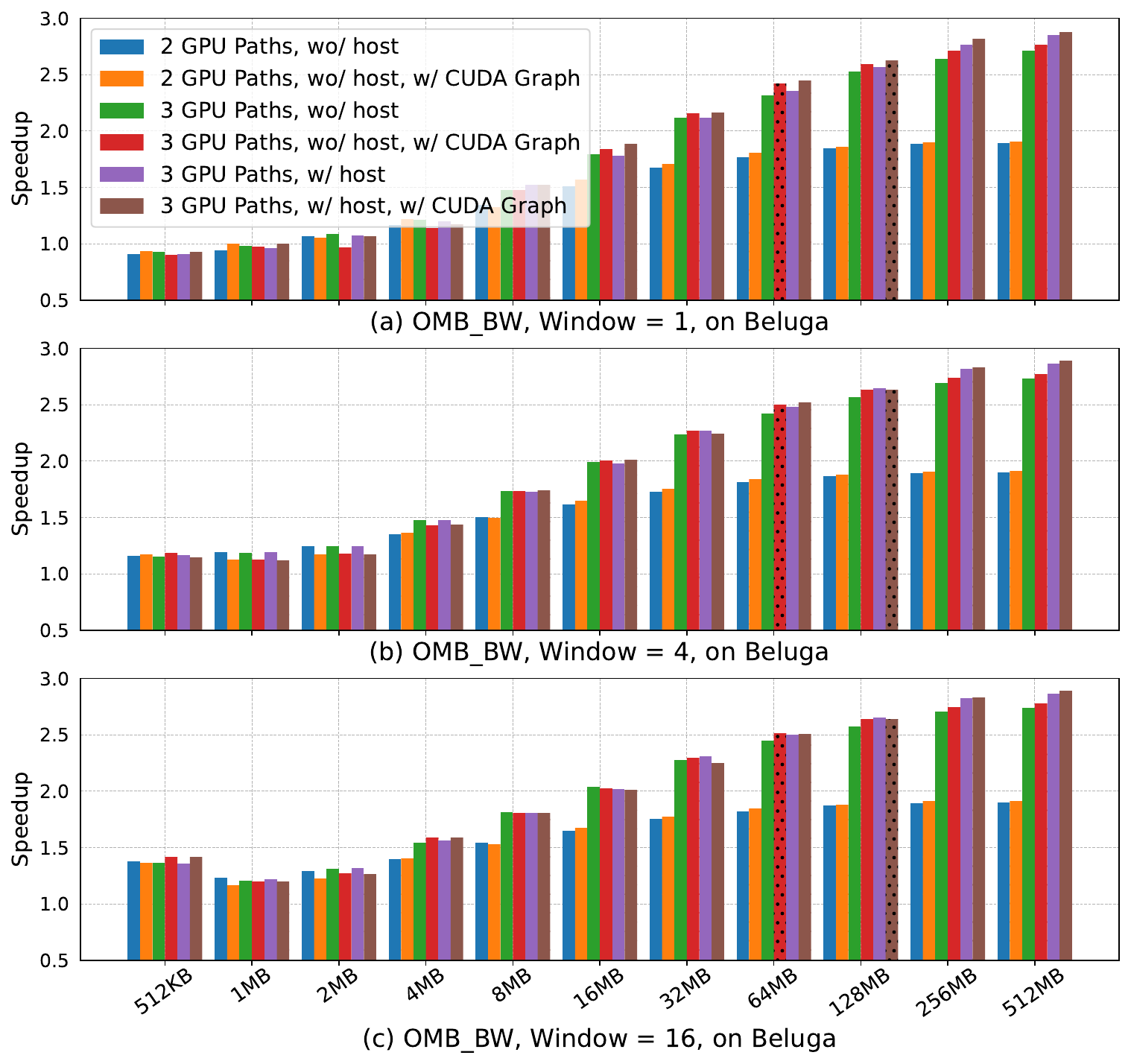}
  \caption{Multi-Path OMB Unidirectional MPI Bandwidth (BW) comparison with and without CUDA Graph, on Beluga}
  \label{cgraph:figs:omb_bw_beluga}
\end{figure}

\begin{enumerate} [leftmargin=0.35cm]
  \item On Beluga, results are consistent with those observed in the \ucx Put test. On Narval, however, we see less improvement for window size one, which suggests that its four \nvlinks do not saturate as quickly as Beluga's two \nvlinks. This pattern is observable in both classic and \cgraph-enhanced versions.
        
  \item For both platforms, bandwidth utilization increases with higher \emph{window} sizes, particularly for message sizes between 8MB and 64MB. This trend is visible in both unidirectional and bidirectional bandwidth plots.
        
  \item For messages larger than 8MB, enabling \cgraph enhances performance even further, especially when the \emph{window} size is increased to 4 or 16. \cgraph allows for overlapping transfers and reduced launch overhead, leading to better performance. This observation suggests that applications that issue multiple outstanding/non-blocking requests (i.e., larger \emph{window} sizes) can benefit more from \cgraph integration. This means that multiple \cgraphs can be launched concurrently, allowing the \cuda runtime to better schedule and overlap operations.
        
  \item In contrast, for messages smaller than 8MB, the number of operations in the \cgraph is small, and the graph launch overhead becomes significant, negating its benefits. This can be seen in the performance drop for message sizes 512MB and 4MB in \Cref{cgraph:figs:omb_bw_beluga}(a), \Cref{cgraph:figs:omb_bw_beluga}(b), and \Cref{cgraph:figs:omb_bw_beluga}(c), where we have only one or two copy nodes in the \cgraph. This pattern is also observable in Narval results and in bidirectional tests.
        
  \item Enabling host paths generally does not improve performance significantly, especially on Narval with window size one (see \Cref{cgraph:figs:omb_bw_narval}(d) and \Cref{cgraph:figs:omb_bibw_narval}(c)). On Narval, the difference in bandwidth between \nvlink and the host-staging path is more prominent than Beluga, and the host path contribution to overall bandwidth is minimal.
        
  \item In bidirectional tests, enabling the host path consistently degrades performance, especially in \cgraph versions. This is because both directions share the same \pcie link to the host, creating contention and bottlenecks. This pattern is more pronounced with higher \emph{window} sizes, as more messages are in-flight. See \Cref{cgraph:figs:omb_bibw_beluga}(c) and \Cref{cgraph:figs:omb_bibw_narval}(c) for instance.
\end{enumerate}

We also observed performance drops for message sizes of 1MB and 2MB, which we suspect are caused by internal algorithm or protocol switches within \ucx or \mpi. This phenomenon is under further investigation.


\subsection{Jacobi Iterative Solver}

\begin{figure}[!t]
  \centering
  \includegraphics[width=0.95\columnwidth]{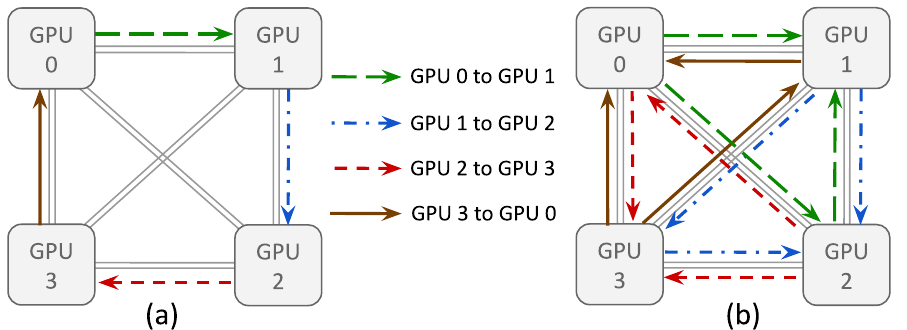}
  \caption{Jacobi communication pattern (a) without multi-path, and (b) with multi-path (two paths per communication) on a four-\gpu node}
  \label{multipath:figs:Multi-GPU-Node-Jacobi}
\end{figure}

To validate the framework in an application scenario, we evaluated the performance of the Jacobi iterative solver using \mpi with one rank per \gpu (a total of four ranks per node). Each rank computes a subregion of the domain and exchanges boundary data with its immediate neighbors using a 2-D halo exchange pattern, as shown in Figure \ref{multipath:figs:Multi-GPU-Node-Jacobi}(a), which is basically a \emph{ring}. Considering the unutilized diagonal links, and the \nvlink's bidirectional feature, $2/3$ of the total available bandwidth is unused in this pattern. Therefore, we can enable multi-path communication for all of these data exchanges, and select the staging \gpus in a way that there is no contention on the \nvlinks. Figure \ref{multipath:figs:Multi-GPU-Node-Jacobi}(b) demonstrates how each communication is split into two paths, and the data is transferred concurrently through the \nvlinks.

\begin{figure}[!t]
  \centering
  \begin{subfigure}[t]{0.80\columnwidth}
    \centering
    \includegraphics[width=\linewidth]{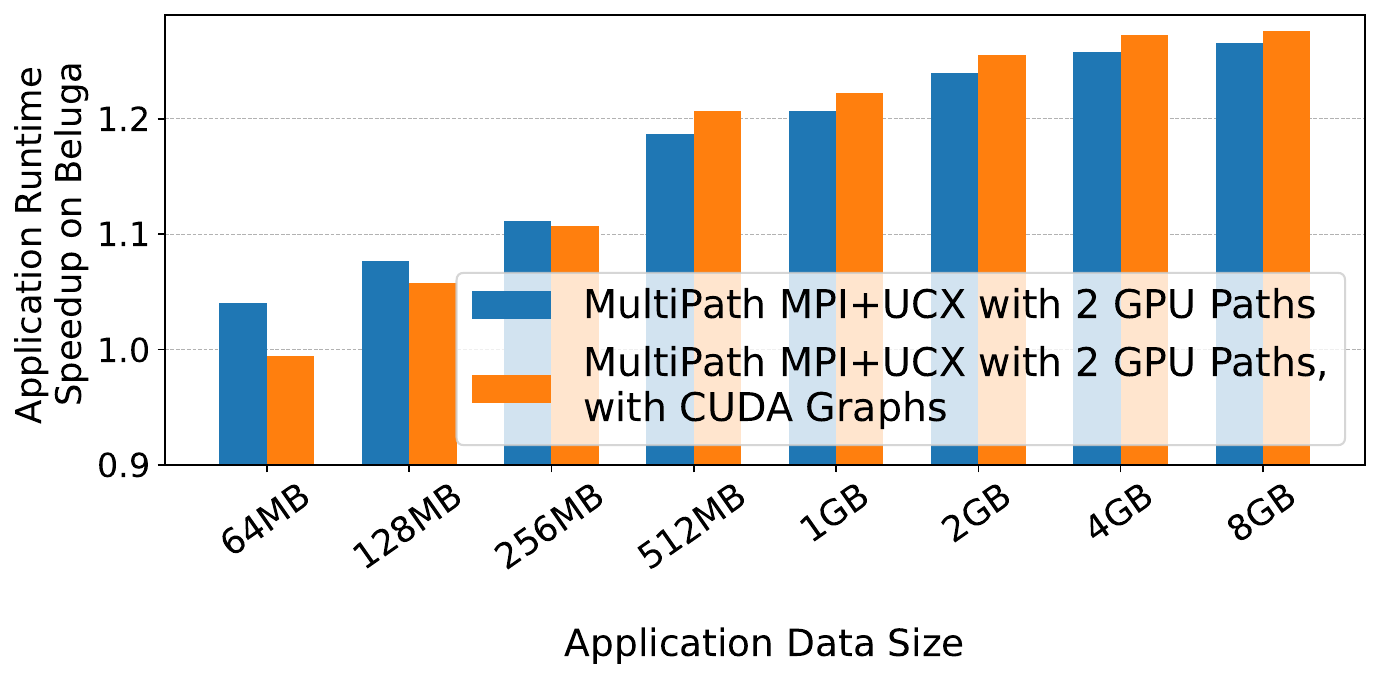}
    \caption{Beluga with two GPU paths}
    \label{fig:jacobi-beluga-gp2}
  \end{subfigure}
  \\
  \begin{subfigure}[t]{0.80\columnwidth}
    \centering
    \includegraphics[width=\linewidth]{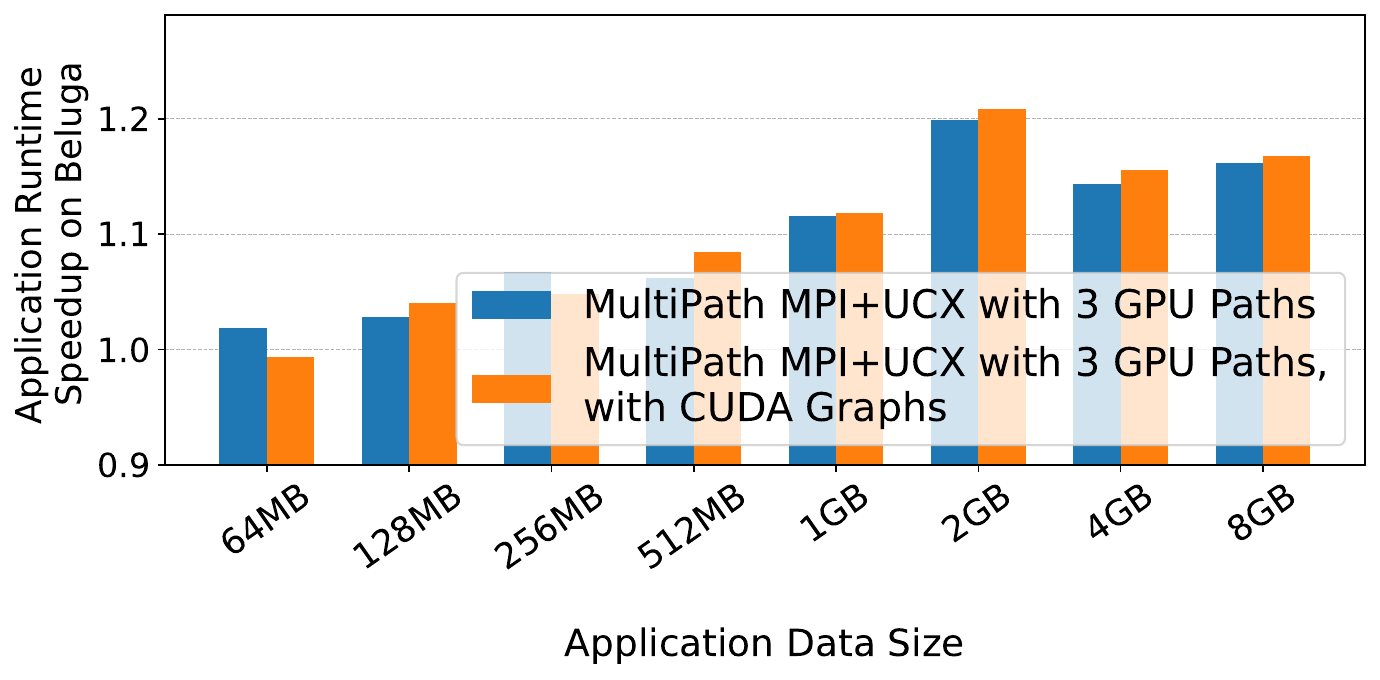}
    \caption{Beluga with three GPU paths}
    \label{fig:jacobi-beluga-gp3}
  \end{subfigure}
  \\
  \begin{subfigure}[t]{0.80\columnwidth}
    \centering
    \includegraphics[width=\linewidth]{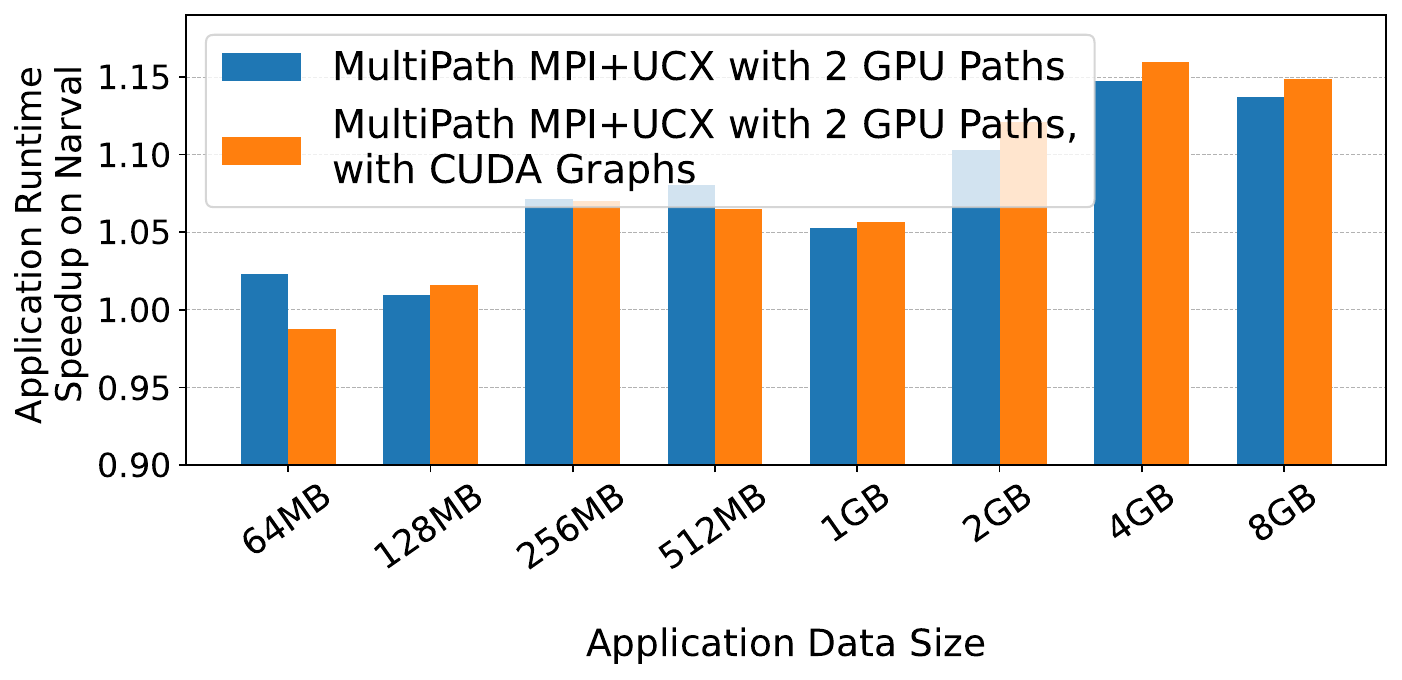}
    \caption{Narval with two GPU paths}
    \label{fig:jacobi-narval-gp2}
  \end{subfigure}
  \\
  \begin{subfigure}[t]{0.80\columnwidth}
    \centering
    \includegraphics[width=\linewidth]{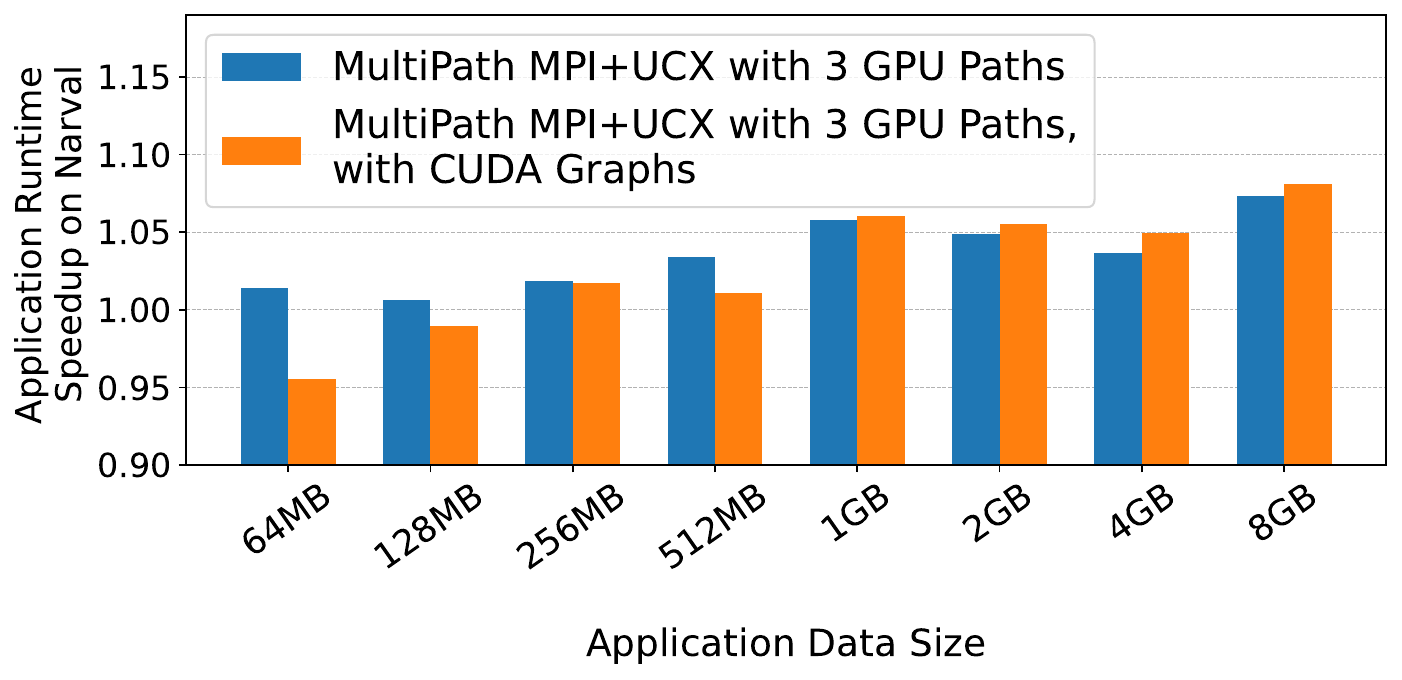}
    \caption{Narval with three GPU paths}
    \label{fig:jacobi-narval-gp3}
  \end{subfigure}
  \caption{Jacobi runtime speedup over default \ucx (\uct::CUDA-IPC) using four \mpi ranks on Beluga and Narval clusters}
  \label{multipath:figs:figure_jacobi_speedup}
\end{figure}

We varied the problem size by fixing the vertical dimension to 8 and increasing the horizontal dimension from $2^{23}$ to $2^{30}$. This means that for the total application data size of 8GB on four \gpus, each rank exchanges 256MB of boundary data with its two neighbors in each iteration. We ran the solver for 1000 iterations and measured the total execution time. We also disabled host staging for this evaluation, as it consistently degraded performance in previous tests. The results, shown in Figure \ref{multipath:figs:figure_jacobi_speedup}, are compared against the default \ucx (\uct::CUDA-IPC) configuration. The key findings indicate:

\begin{enumerate} [leftmargin=0.35cm]
  \item Using two concurrent paths per communication improves application runtime by up to 1.26$\times$ and 1.15$\times$ on Beluga and Narval, respectively.
  \item Integration of \cgraphs results in further performance gains, achieving up to 1.28$\times$ and 1.16$\times$ speedup on Beluga and Narval, respectively, for larger problem sizes.
  \item As discussed earlier, the benefits of \cgraphs are more pronounced for larger messages due to the increased number of operations in the graph, which helps amortize the launch overhead. For smaller problem sizes, the overhead of \cgraph creation and launch can outweigh its benefits, leading to marginal or no improvement: for application sizes smaller than 1GB (message size $=$ 32MB), in which the number of operations in the \cgraph is insufficient to amortize the overhead. This is also consistent with our observations in the bidirectional micro-benchmarks (\Cref{multipath:results.omb}).
  \item Even with three \gpu paths per transfer, potentially incurring additional contention, our framework still improves the runtime. Although the gains are smaller than with two paths, we still observe improvements of up to 1.2$\times$ and 1.08$\times$ on Beluga and Narval, respectively. This suggests that the benefits of increased concurrency do not outweigh the contention costs in this scenario, and two paths may be the optimal choice.
  \item It is important to note that in this application we are only using the \cgraphs for enabling multi-path transfers, and none of the other operations (e.g., computation kernels) are included in the \cgraph. The communication patterns created and launched in this test map to a graph which has a very small number of vertices along each path (only two nodes per chunk per path). This limits the potential for optimization from \cgraph utilization. These results show that in scenarios with more complex communication patterns, the advantages of \cgraphs could be more significant, especially when kernel executions are involved.
  \item Importantly, Jacobi's numerical convergence remains unaffected, demonstrating the correctness of our pipelined and concurrent transfer mechanisms.
\end{enumerate}

\subsection{CUDA Graph Overhead Analysis}

In this section, we characterize the overheads associated with various \cgraph operations in our multi-path framework, including creation, construction, instantiation, and launch. This analysis helps us understand the trade-offs involved in using \cgraphs, particularly in terms of latency reduction and overhead minimization for different message sizes and number of nodes in the graph. These insights can guide users in deciding when and how to effectively utilize \cgraphs in their applications.

We collected timing data during the \omb \code{Latency} micro-b\-e\-n\-c\-h\-m\-a\-rk on Narval for messages between 2MB and 512MB (note that multi-pathing starts at 2MB with only two nodes in the graph). In more detail, these are the various phases:

\begin{enumerate} [leftmargin=0.35cm]
  \item \textbf{Creation} involves allocation and initialization of the \cgraph object (\code{cudaGraph\_t}) during the first time execution. For subsequent executions, this step is skipped, and the existing \cgraph object is fetched from a hash table and reused.
  \item \textbf{Construction} includes defining and creating the graph nodes, which denotes the data movements, as well as their inter-dep\-e\-n\-d\-e\-n\-c\-i\-e\-s. We used the \cgraph explicit \api to construct the \cgraph design and update the existing \cgraph object. Similar to \emph{creation}, this step is only performed during the first execution, and the constructed graph is reused in subsequent executions.
  \item \textbf{Instantiation} is a one-time expensive operation that allocates resources for the \cgraph execution and optimizing it. After this step, the object \code{cudaGraphExec\_t} is created, which can be launched multiple times. This step is synchronous, and its overhead can vary based on the complexity of the graph.
  \item \textbf{Launch} refers to the initiation of the actual execution of the \cgraph on the \gpu. Similar to a \cuda kernel launch, this process is asynchronous and expensive, and its overhead might vary depending on the complexity of the graph.
\end{enumerate}

\begin{figure*}[t!]
  \centering
  \begin{subfigure}[t]{0.92\columnwidth}
    \centering
    \includegraphics[width=\linewidth]{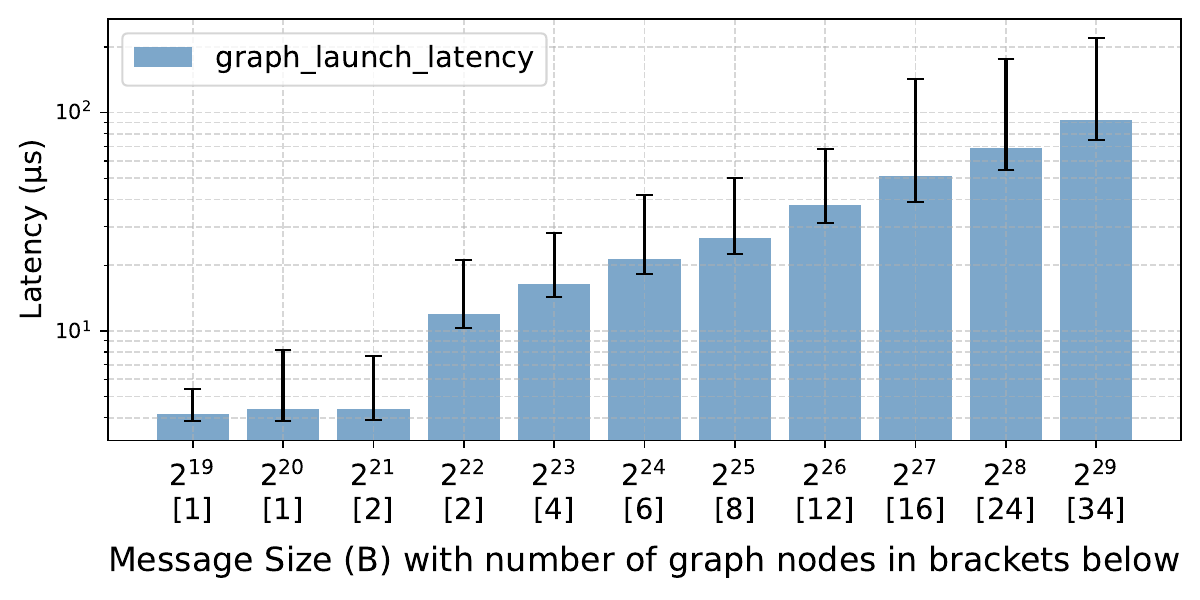}
    \caption{Overhead of the subsequent CUDA Graph launches for various node counts.}
    \label{fig:omb-lat-graph-launch-latency-bar-rest}
  \end{subfigure}
  \begin{subfigure}[t]{0.92\columnwidth}
    \centering
    \includegraphics[width=\linewidth]{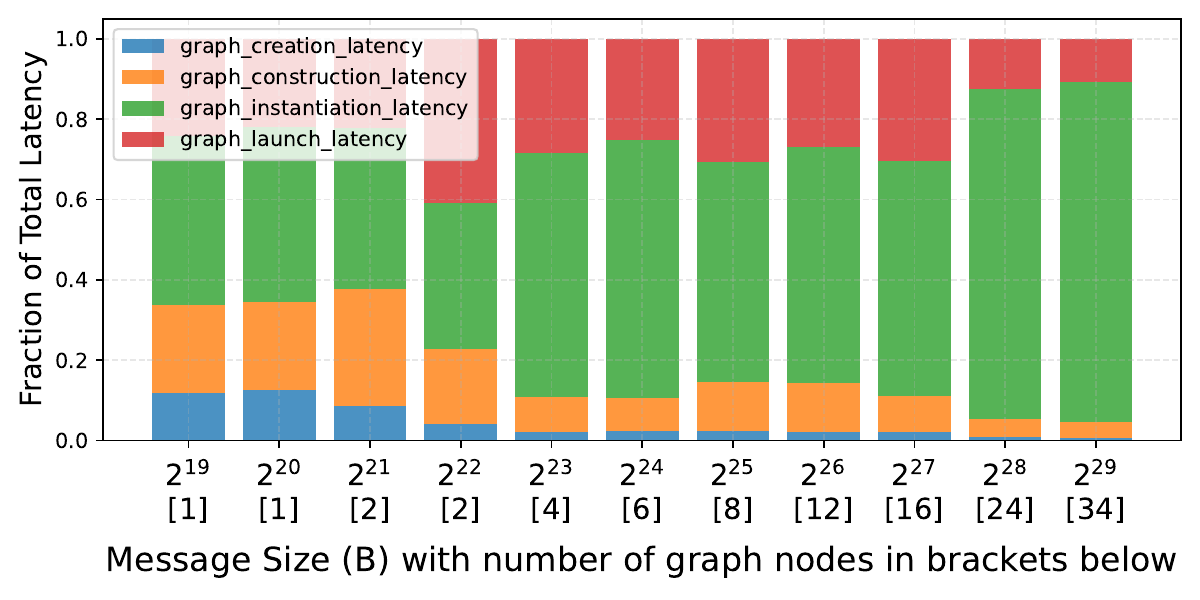}
    \caption{Contribution of each graph phase to the total latency (first iteration).}
    \label{fig:omb-lat-graph-phase-fraction-first}
  \end{subfigure}
  \begin{subfigure}[t]{0.92\columnwidth}
    \centering
    \includegraphics[width=\linewidth]{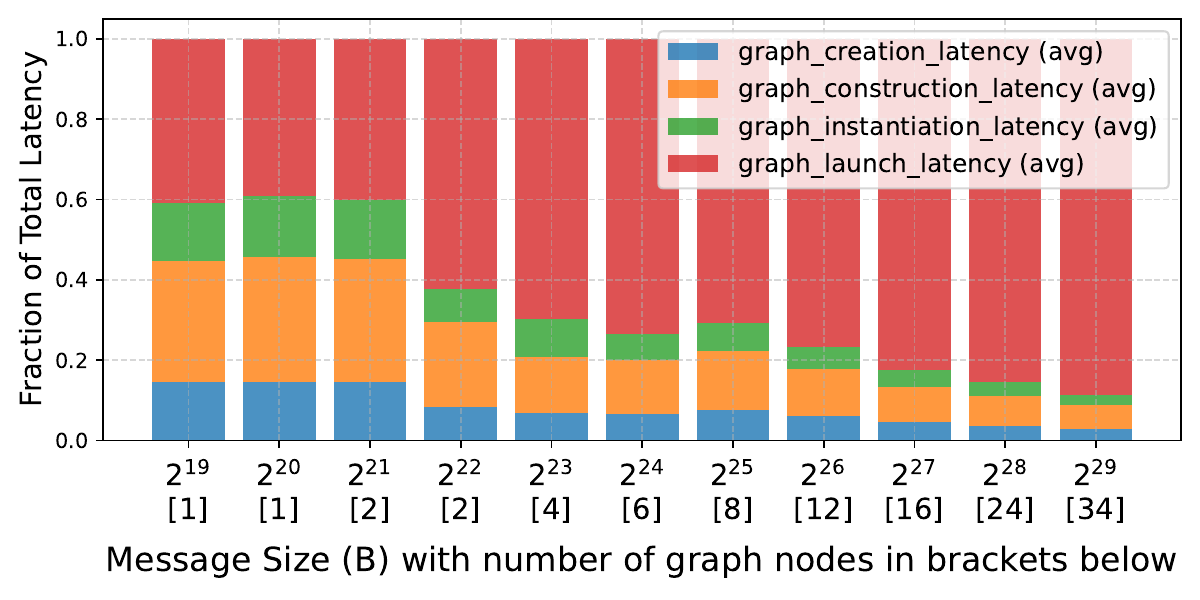}
    \caption{Contribution of each graph phase to the total latency (subsequent iterations).}
    \label{fig:omb-lat-graph-phase-fraction-rest}
  \end{subfigure}
  \begin{subfigure}[t]{0.92\columnwidth}
    \centering
    \includegraphics[width=\linewidth]{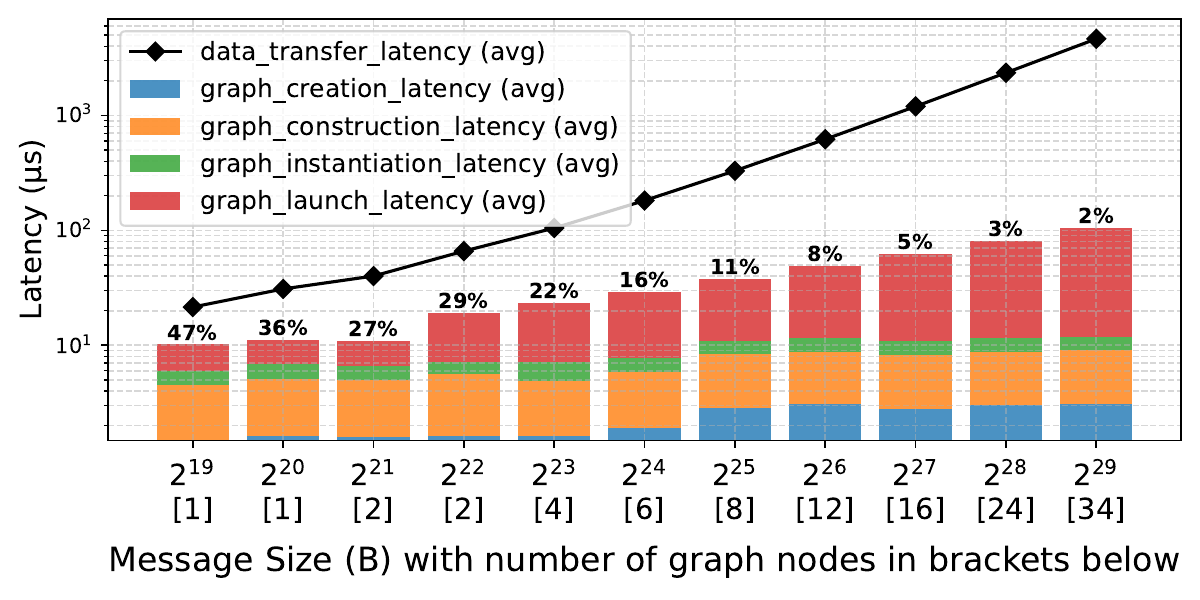}
    \caption{Comparison of subsequent CUDA Graph phases with the CUDA Graph enabled multi-path transfer.}
    \label{fig:omb-lat-graph-box-bar-rest}
  \end{subfigure}
  \caption{Measurement of various CUDA Graph operations during OMB Latency benchmark on Narval for dual-path communication from message sizes 2MB to 512MB. Note that multi-path is enabled from 2MB.}
  \label{cgraph:figs:omb_lat_overheads_analysis_b}
\end{figure*}

\Cref{cgraph:figs:omb_lat_overheads_analysis_a} and \Cref{cgraph:figs:omb_lat_overheads_analysis_b} demonstrate how the overhead of various \cgraph operations changes according to the number of nodes within the graph when they are called for the first/subsequent times. Analyzing the results, we observe the following:

\begin{figure*}[t!]
  \centering
  \begin{subfigure}[t]{0.92\columnwidth}
    \centering
    \includegraphics[width=\linewidth]{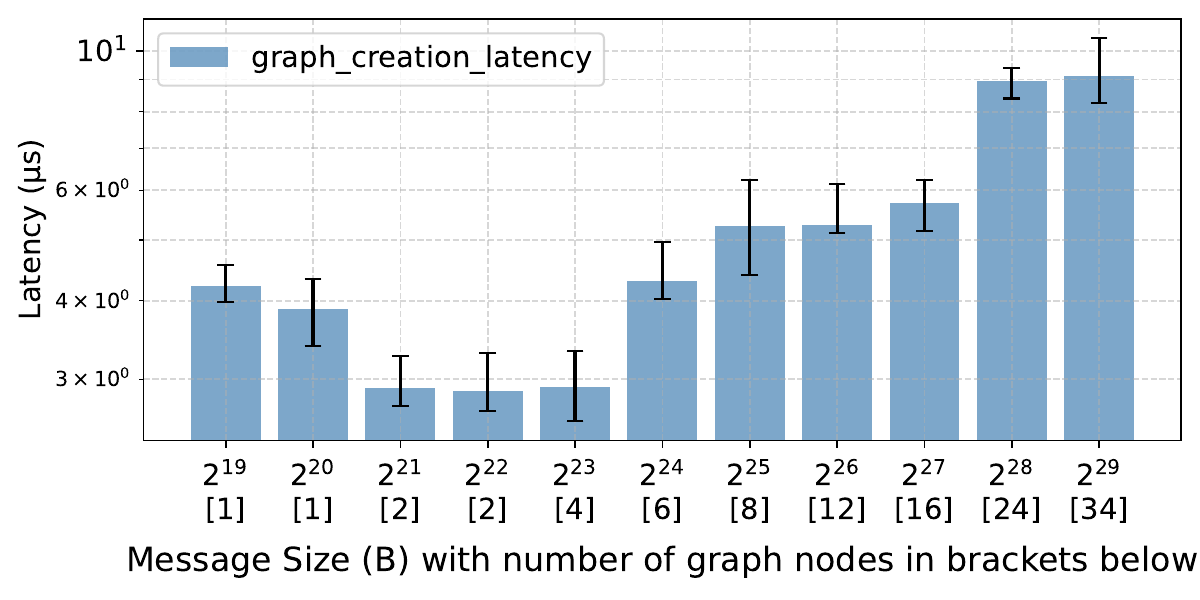}
    \caption{Overhead of the first CUDA Graph creation.}
    \label{fig:omb-lat-graph-creation-latency-bar-first}
  \end{subfigure}
  \begin{subfigure}[t]{0.92\columnwidth}
    \centering
    \includegraphics[width=\linewidth]{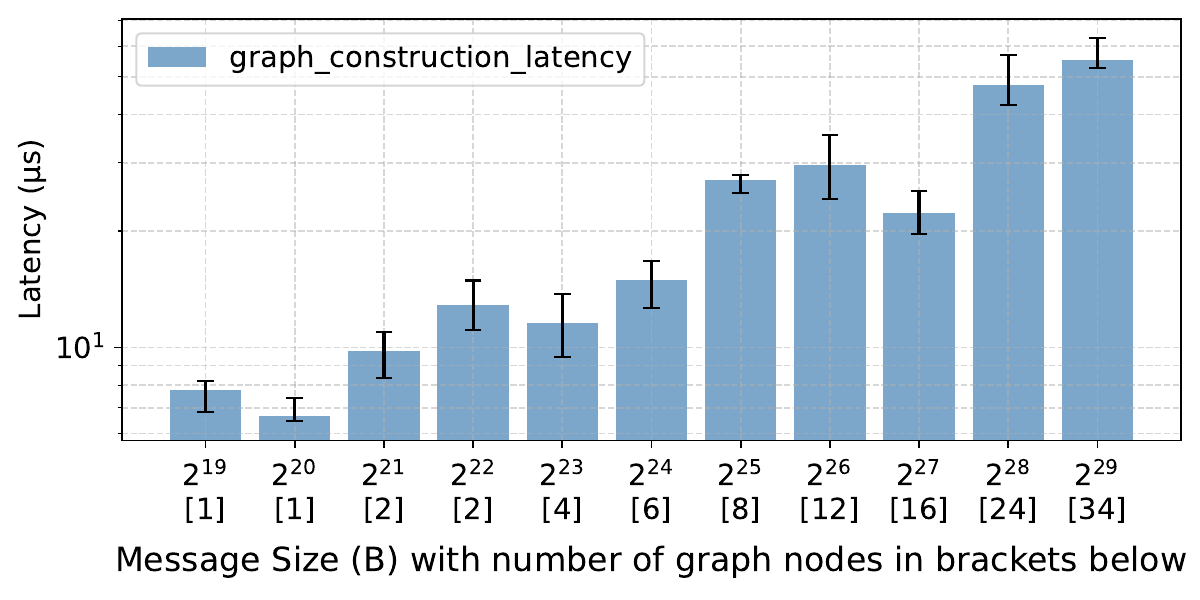}
    \caption{Overhead of the first CUDA Graph construction.}
    \label{fig:omb-lat-graph-construction-latency-bar-first}
  \end{subfigure}
  \begin{subfigure}[t]{0.92\columnwidth}
    \centering
    \includegraphics[width=\linewidth]{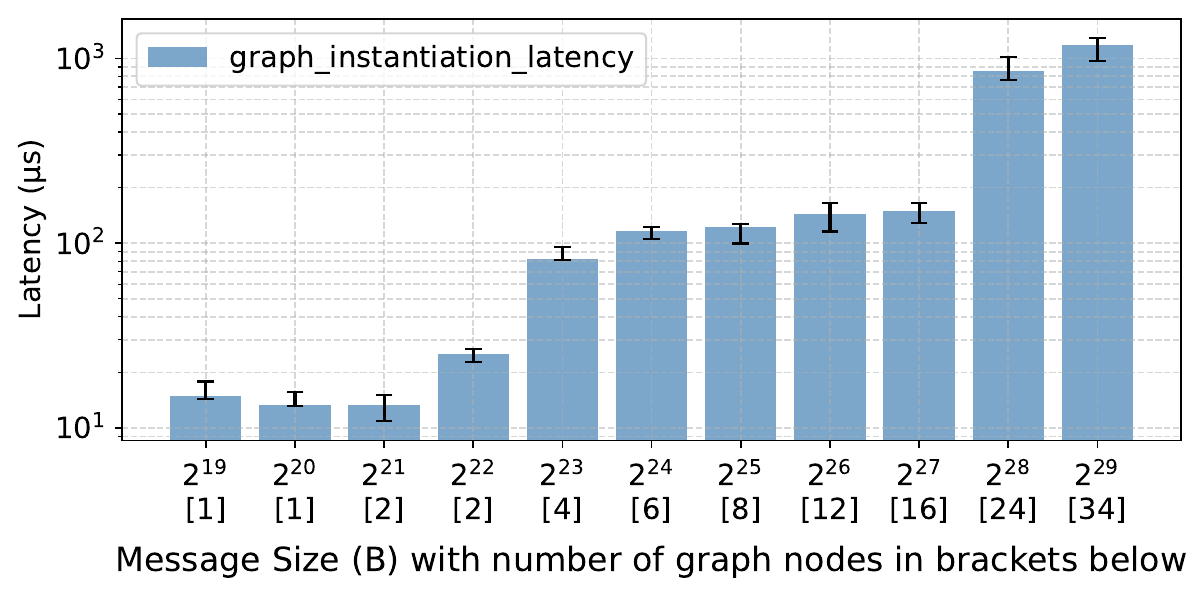}
    \caption{Overhead of the first CUDA Graph instantiation.}
    \label{fig:omb-lat-graph-instantiation-latency-bar-first}
  \end{subfigure}
  \begin{subfigure}[t]{0.92\columnwidth}
    \centering
    \includegraphics[width=\linewidth]{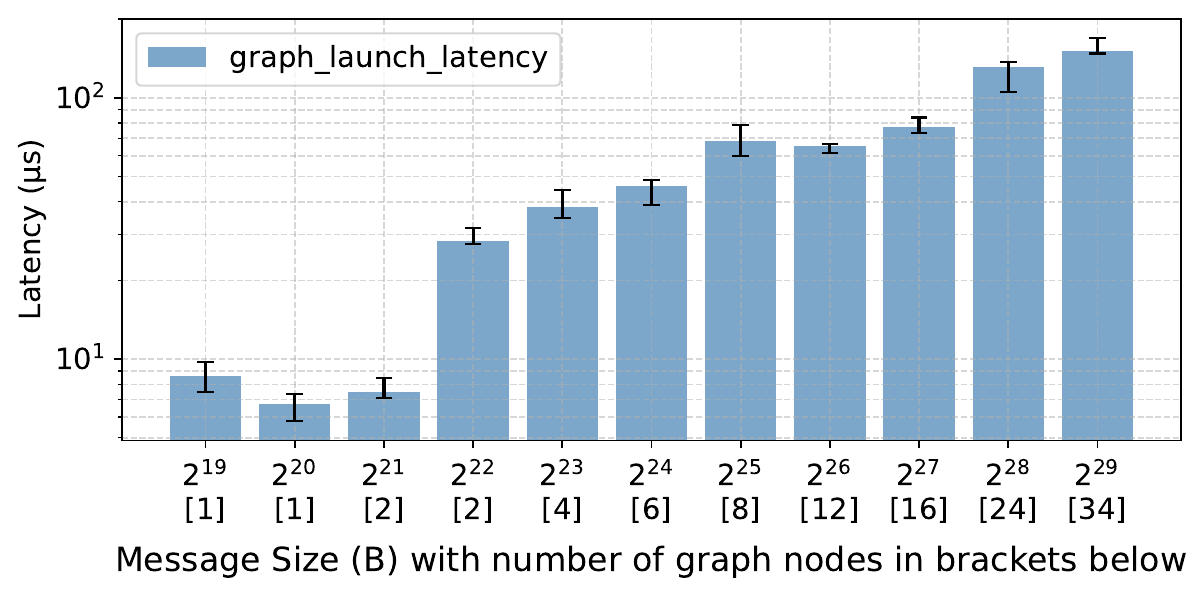}
    \caption{Overhead of the first CUDA Graph launch.}
    \label{fig:omb-lat-graph-launch-latency-bar-first}
  \end{subfigure}
  \caption{Measurement of various CUDA Graph operations during the first iteration of OMB Latency benchmark on Narval for dual-path communication from message sizes 2MB to 512MB. Note that multi-path is enabled from the 2MB.}
  \label{cgraph:figs:omb_lat_overheads_analysis_a}
\end{figure*}

\begin{itemize}[leftmargin=0.35cm]
  \item \textbf{Observation 1:} In all the graph phases during the first iteration, the overhead increases with the number of nodes. See \Cref{cgraph:figs:omb_lat_overheads_analysis_a}(a), \Cref{cgraph:figs:omb_lat_overheads_analysis_a}(b), \Cref{cgraph:figs:omb_lat_overheads_analysis_a}(c), and \Cref{cgraph:figs:omb_lat_overheads_analysis_a}(d) for \emph{creation}, \emph{construction}, \emph{instantiation}, and \emph{launch}, respectively. This is expected, as more nodes imply more operations and dependencies to manage. Notably, \emph{instantiation} depicts the highest overhead, reaching up to 3ms for 34 nodes (512MB message size with dual-path communication).
  \item \textbf{Observation 2:} During the subsequent calls, the overhead stays constant and low for all the cases, except for the launch phase. \Cref{cgraph:figs:omb_lat_overheads_analysis_b}(a) shows how the launch overhead increases with the number of nodes (figures of other phases are not provided here for brevity).
  \item \textbf{Observation 3:} Contribution of each graph phase to the total latency varies significantly between the first and subsequent iterations. While \emph{instantiation} is the most dominant operation during the first iteration, the \emph{launch} phase becomes more significant in subsequent iterations. Compare \Cref{cgraph:figs:omb_lat_overheads_analysis_b}(b) and \Cref{cgraph:figs:omb_lat_overheads_analysis_b}(c).
  \item \textbf{Observation 4:} The overall latency benefits from the multi-path optimization become more pronounced as the message size increases. Although the overhead of \cgraph operations increases with node count (and message size), the relative impact on latency decreases (See \Cref{cgraph:figs:omb_lat_overheads_analysis_b}(d)). These results suggest that \cgraphs provide scalable performance improvements as the complexity of the communication pattern increases.
\end{itemize}

Overall, these results suggest that while \cgraph operations introduce some overhead, especially during the first iteration, the benefits of multi-path communication outweigh these costs for larger messages. As we saw, the overhead of \emph{creation}, \emph{construction}, and \emph{instantiation} is amortized over multiple launches, making them negligible for long-running applications. This is why libraries such as \ucx can benefit from \cgraphs, as they can create and instantiate the graphs once and reuse them multiple times for recurring communication patterns.

\section{Conclusions and Future Work}
\label{multipath:conclusion}

We proposed a \cgraph-based multi-path communication framework implemented within the \ucx library that reuses communication workflows to minimize launch overhead and improve intra-node communication efficiency. By leveraging available communication channels, our 2-D pipelining engine scatters \ptp communication across both \nvlink and \pcie channels to maximize communication bandwidth between intra-node \gpus. We also provide end users with tuning capabilities for scheduling parameters, ensuring adaptability to diverse communication patterns and hardware configurations.

Our experimental evaluation on a four-\gpu node demonstrates that this approach achieves up to a $2.95\times$ increase in \gpu-to-\gpu \omb \mpi bandwidth test compared to \ucx single-path method for very large messages ($\geq 32MB$). We also observed that besides utilizing unused interconnects, harnessing \nvlink's bidirectional features will also improve communication performance. However, the lack of the same feature in \pcie communication channels may lead to contention, and consequently, performance degradation. Finally, we showed that our multi-path communication framework can improve the performance of the Jacobi iterative solver by up to $1.28\times$.

We showed that the benefits of \cgraphs were marginal relative to the non-\cgraph multi-path framework, due to the limited number of operations in the graph. Our results suggest that in scenarios with more complex communication patterns, the advantages of \cgraphs could be more significant, especially when kernel launches are involved, and when the overhead of kernel/communication launches are considerable compared to the actual execution time of these operations.

Furthermore, we discussed the overheads associated with various \cgraph operations in the context of multi-path communication, including creation, construction, instantiation, and launch. Our analysis indicates that while these operations introduce some overhead, particularly during the first iteration, the benefits of multi-path communication outweigh these costs for larger messages, and the overheads are amortized over multiple launches, making them negligible for long-running applications.

While our approach can improve the communication bandwidth, the performance gain is dependent on the concurrent communication pattern. This is why we believe that a more adaptive approach could yield better results. A possible future direction is to dynamically adapt the communication pattern based on both the application's communication pattern and the hardware configuration. Moreover, we plan to extend our framework to optimize our \cgraphs engine to minimize their creation and instantiation time, applying smarter caching mechanisms, and dynamically adjusting the \cgraphs structure and parameters through their explicit \api. 

Another interesting future direction is to explore the integration of our multi-path communication framework with collective communication operations, such as Allreduce, and Alltoall. By leveraging the benefits of multi-path communication in designing collective operations, we can potentially achieve significant performance improvements for applications that rely heavily on these operations. 

Finally, we plan to empirically validate the scalability of our framework on 8+ \gpu nodes and multi-node architectures, including \nvswitch topologies and DGX systems. We believe that our approach can provide substantial benefits in these topologies, where multiple high-bandwidth links are available between \gpus.

\section{Acknowledgments}

This research was supported in part by the Natural Sciences and Engineering Research Council of Canada and Digital Research Alliance of Canada. Computations were performed on Beluga and Narval with support from Calcul Québec (calculquebec.ca).

\bibliographystyle{ACM-Reference-Format}
\bibliography{References}


\end{document}